\documentclass[%
aps,prb,superscriptaddress,twocolumn,floatfix,10pt]{revtex4-2}
\usepackage[utf8]{inputenc}
\usepackage{amsmath,amssymb}
\usepackage{empheq}
\usepackage{lipsum}
\usepackage{mathtools,cuted}
\usepackage{esvect}
\usepackage{multirow}
\usepackage{xcolor}
\usepackage{soul}
\usepackage[export]{adjustbox}
\usepackage{graphicx}
\usepackage{dcolumn}
\usepackage{bm}
\usepackage[colorlinks=true,citecolor=blue]{hyperref}
\usepackage{nameref}
\usepackage{physics}
\usepackage[mathlines]{lineno}
\usepackage[capitalise]{cleveref}
\usepackage{bbm}
\usepackage{orcidlink}
\usepackage{comment}
\usepackage{enumitem}
\usepackage{tikz}
\usetikzlibrary{automata, positioning, arrows.meta}
\usetikzlibrary{calc}
\usepackage{bbold}
\usepackage{overpic}

\newcommand{\MK} [1] {\color{orange}MK: {#1}\color{black}\normalsize}
\newcommand{\AM} [1] {\color{olive}\textbf{AM: #1}\color{black}\normalsize}

\begin{document}

\title{One-dimensional quasicrystals with tensor-network finite-state automata}

\author{Milla Kolehmainen \,\orcidlink{0009-0005-6953-8890}}
 \affiliation{Department of Applied Physics, Aalto University, 02150 Espoo, Finland}
\author{Jose L. Lado\,\orcidlink{0000-0002-9916-1589}}
 \affiliation{Department of Applied Physics, Aalto University, 02150 Espoo, Finland}
 \author{Anouar Moustaj\,\orcidlink{0000-0002-9844-2987}}
 \affiliation{Department of Applied Physics, Aalto University, 02150 Espoo, Finland}

\date{\today}

\begin{abstract}
Quasicrystals occupy a distinctive position between the translational order of crystals and the disordered amorphous matter. Simulating this physics has remained challenging, since quasiperiodic structures lack the translational symmetry exploited for crystals and generally require costly diagonalization of large finite approximants.
Quasicrystalline order admits two equivalent descriptions, a cut-and-project scheme from a higher-dimensional periodic crystal, and a discrete set of substitution rules acting on a finite alphabet. We show that the latter, written in a numeration system adapted to the substitution, defines a deterministic finite automaton with output, the digits of a site index are fed, and the automaton returns the letter occupying that site. We further exploit another equivalence to a different construction, the transition matrices are exactly the tensors of a matrix product state, whose bond dimension is the number of automaton states and is independent of system size. This allows efficient representation of extremely large tight-binding Hamiltonians in the tensor-train language, thereby yielding an exact matrix product operator for the quasicrystal Hamiltonian at any system size. We show how this framework works for two families of one-dimensional quasicrystals, the metallic-mean and $k$-bonacci families and we explicitly construct the Fibonacci, silver-mean, and Tribonacci quasicrystals. By leveraging efficient tensor-network compression and the kernel polynomial method, we compute spectral densities for chains with more than $10^9$ sites and directly resolve the hierarchical structure of the spectrum.
\end{abstract}
	
\maketitle

\section{Introduction}\label{sec:Intro}

Quasiperiodic systems possess long-range order without translational symmetry, which places them in a middle ground between the perfect periodicity of crystals and the disorder of amorphous materials. Quasicrystals are a well-known class of aperiodic systems, which, just like periodic crystals, possess a pure-point diffraction spectrum \cite{Shechtman1984MetallicSymmetry}. Their unique structural order gives rise to several exotic physical phenomena, including fractal energy spectra, critical wavefunctions, and anomalous transport \cite{Kohmoto1984CantorMap,Kohmoto1987CriticalModel,Niu1986Renormalization-GroupSystems,Piechon1996AnomalousChains}. More recently, interest in these systems occasionally reemerges as they offer an unconventional platform to study anomalous behavior for several physical phenomena, such as localization behavior \cite{Jagannathan2019NonmonotonicQuasicrystal,Moustaj2021EffectsQuasicrystal,Mace2019Many-bodyChain}, topological phases \cite{Kraus2012TopologicalQuasicrystals,Verbin2015TopologicalQuasicrystal,Rai2021BulkQuasicrystal,Yoshii2021TopologicalIndex,Moustaj2025AnomalousInsulators}, superconductivity \cite{Rai2019ProximityRing,Sun2024EnhancementChain,Sardinero2026JosephsonStates}, and beyond also in mathematics, with the recently constructed hat tiling in 2D \cite{Smith2024AnMonotile,Schirmann2024PhysicalModes,RocheCarrasco2025FamilyTensor}. 
Quasiperiodicity is typically introduced via the lattice geometry or via aperiodic modulation of the on-site potentials or the hopping amplitudes in a tight-binding model. However, the absence of translational symmetries removes standard symmetry simplifications, making large-scale simulations computationally challenging. Nevertheless, the deterministic nature of quasiperiodic order suggests that these systems may remain accessible for efficient numerical modeling given an appropriate compression method.

In quantum many-body physics, tensor network methods are a powerful tool to model systems with exponentially large Hilbert spaces, providing accurate solutions to many paradigmatic problems, including Hubbard and Heisenberg models \cite{White1992DensityGroups,Cirac2021MatrixTheorems,Banuls2023TensorMap,Orus2014AStates,Schollwock2011TheStates,Stoudenmire2012StudyingGroup,Fishman2022TheCalculations,Fishman2022CodebaseITensor,Zhou2020WhatComputers,Huggins2019TowardsNetworks,Chan2016MatrixAlgorithms,Orus2019TensorSystems}. 
Their success relies on the fact that the physically relevant corner of this exponentially large Hilbert space is typically characterized by limited entanglement between subsystems, allowing the many-body wavefunction to be decomposed into a network of smaller, locally connected tensors rather than a single object of exponential size.
Tensor networks work by representing multidimensional tensors in a compressed form, such that the exponential number of elements required to specify the full tensor is approximated by the number of cores in the network, weighted by the bond dimensions connecting them.
More recently, quantics tensor cross interpolation (QTCI) \cite{Oseledets2010TT-crossArrays,Oseledets2011Tensor-TrainDecomposition,Ritter2024QuanticsFunctions,Jeannin2024Cross-extrapolationRegime,NunezFernandez2025LearningLibraries,Ritter2022QuanticsTCI.jl,Ritter2022TensorCrossInterpolation.jl} has broadened the applicability of tensor networks by enabling the representation of an arbitrary function as a matrix product state (MPS). Building on tensor representations, tensor networks have demonstrated speed ups and increases in accuracy in many different fields such as machine learning \cite{Stoudenmire2016SupervisedNetworks,Dilip2022DataLearning,Han2018UnsupervisedStates}, computational chemistry \cite{Jolly2025TensorizedChemistry} and many other computational problems \cite{Waintal2026WhoTechniques,Shinaoka2023MultiscaleTrains,Erpenbeck2023TensorModels,NunezFernandez2022LearningTrains,Takahashi2025CompactnessPropagators,Jeannin2025ComprehensiveBlockade,Rohshap2025Two-particleEquations}.  Furthermore, a variety of tensor-network methodologies based on the QTCI encoding has recently been shown able to
tackle a variety of ultra-large tight-binding problems \cite{Antao2026TensorApplications,Fumega2025CorrelatedAlgorithm,Sun2025Self-consistentSites,Antao2025TensorMosaics,Moustaj2026Tensor-networkSites,Moustaj2026TensorSystems,Sun2026Real-spaceNetworks}. 
However, the QTCI encoding fails to compress the purely quasiperiodic substitution rules that generate quasicrystals. The reason for this is that it is impossible to generate a finite set of Fibonacci rules in the binary number system employed by the QTCI encoding.

Here we propose in this manuscript an encoding strategy that accounts for the underlying geometry of quasicrystals
generated by an encoding substitution, not tractable with a conventional quantics encoding. 
The key observation is that MPSs provide representations of deterministic finite automata with output (DFAOs) \cite{Allouche2003FiniteComputation,Crosswhite2008FiniteAlgorithms,Li2024ConnectingLearning}. Discrete substitution sequences can, in turn, be generated by finite automata whenever they admit a finite-state description in a suitable numeration system. The quantics encoding is purely binary and therefore naturally suited to the efficient compression of $2$-automatic sequences. Many aperiodic sequences, however, are not $2$-automatic, making a binary representation poorly adapted to their underlying structure. 
More generally, a $k$-automatic sequence can be efficiently represented using a base-$k$ numeration system. On the other hand, quasicrystalline sequences, forming a special class of aperiodic sequences, do not admit finite-state representations in conventional integer bases. Paradigmatic examples such as the Fibonacci quasicrystal nevertheless become finite-state when expressed in numeration systems adapted to their substitution rules. 
In particular, the Fibonacci sequence can be generated by a DFAO in the Fibonacci numeration system, a binary positional representation subject to constraints on the allowed digit strings. Therefore by encoding the sequence in this MPS structure, one can cheaply generate a highly compressed representation of the Fibonacci sequence, which in turn can be encoded in the tight-binding parameters of the tensor-network representation of the Hamiltonian matrix product operator (MPO). This allows one to construct Hamiltonians of sizes beyond $N\sim10^9$ sites at a cheap memory cost $\mathcal{O}(\log N)$, and resolve their spectral features in real-space by using a tensor-network kernel polynomial method.

The paper is organized as follows. In \cref{Sec: Aperiodic Sequences MPS} we show that aperiodic sequences generated by substitution rules become DFAOs once the site index is written in an adapted numeration system, and that these automata are encoded exactly as MPSs. We develop the argument in detail for the Fibonacci chain before generalizing it to the metallic-mean and $k$-bonacci families. In \cref{Sec: QC Ham MPO} we apply this formalism to construct Hamiltonian MPOs for quasicrystalline systems, and demonstrate the method by computing spectral densities resolved in real space for systems beyond $10^9$ sites. Finally, \cref{Sec: Conclusion} collects our concluding remarks.

\section{Aperiodic Sequences as Matrix Product States}\label{Sec: Aperiodic Sequences MPS}

Our method relies on representing aperiodic sequences as an MPS, which are further used as building blocks to represent aperiodic Hamiltonians as MPOs. To enable efficient simulation of an aperiodic system its MPO Hamiltonian, and therefore also its building blocks, should have sufficiently low bond dimensions. The maximum bond dimension of a tensor network controls the complexity and amount of information that can be transferred between local tensors. The crucial step is to find for each aperiodic sequence an MPS encoding in which it can be represented with minimal information. One way to generate aperiodic sequences is to use deterministic substitution rules. These can in turn be represented as DFAOs, for which it has been shown that the maximum bond dimension in the MPS encoding is at most the number of states in the DFAO \cite{Li2024ConnectingLearning}.

\subsection{Aperiodic Systems through Substitution Rules as Deterministic Finite Automaton with Output}\label{Sec: Sub 1.1}

A substitution sequence is a sequence that is generated by repeatedly applying a substitution rule to an initial seed. For example, Fibonacci words are characterized by a substitution rule $\mu$ which acts on a two-letter alphabet $\Sigma=\{A, B\}$ as follows: $\mu(A)=AB$ and $\mu(B)=A$. The Fibonacci word is obtained as a limit of repeated applications of $\mu$ to the initial seed $A$, and it is aperiodic \cite{Jagannathan2021TheMultifractality}.

In particular, we focus on substitution sequences that are both aperiodic and automatic. An automatic sequence is an infinite sequence over a finite alphabet that is generated by a DFAO \cite{Allouche2003FiniteComputation}. Formally, a DFAO is a 6-tuple $M = (S,\Sigma,\delta,s_0,\Omega, \tau)$, where $S$ denotes a finite set of states, $\Sigma$ an input alphabet, $\delta:S\times\Sigma \to S$ a transition map, $s_0\in S$ an initial state, $\Omega$ an output set and $\tau : S \to \Omega$ an output map. A sequence is called $k$-automatic if a DFAO with an input alphabet $\Sigma_k=\{0,1,\dots,k-1\}$ can determine the $n$th symbol of the sequence by reading the base-$k$ representation of $n$. Some well-known examples of 2-automatic aperiodic sequences are Thue-Morse \cite{Allouche1999TheSequence,Baake2019ScalingMeasure} and Rudin-Shapiro \cite{Rudin1959SomeCoefficients,Dulea1992Trace-mapModel} sequences. 

The structure of a finite automaton can be naturally translated into an MPS representation, as will be discussed in \cref{Sec: Sub 1.2}. However, Sturmian words \cite{Lothaire2002SturmianWords}, which represent one of the simplest classes of one-dimensional quasicrystals, are not $k$-automatic for any integer $k\ge2$ \cite{Rampersad2018CommonSequences}.  Nevertheless, for certain Sturmian words, automaton-like structures can still be exploited by extending the conventional notion of automaticity \cite{Allouche2003FiniteComputation,Frougny2010NumberAutomata}. We shall refer to sequences generated by such structures as \textit{generalized} automatic sequences. 

Let us consider the Fibonacci word as an example. According to Zeckendorf's Theorem \cite{Zeckendorf1972RepresentationsLucas}, every positive integer $n$ can be uniquely represented as a sum of non-consecutive Fibonacci numbers as $n = \sum_{i=0}^{L-1}\sigma_{L-i}F_{i+2}$. As an analogue of a base-$k$ representation, the Fibonacci representation of $n$ can be written as $n=(\sigma_1\sigma_2\cdots\sigma_L)_F$, where $\sigma_i\in\{0,1\}$ and $\sigma_i\sigma_{i+1}=0$ for all $i<L$. The Fibonacci word is called Fibonacci-automatic since its $n$th symbol can be determined from the Fibonacci representation of $n$ \cite{Frougny2010NumberAutomata,Mousavi2016DecisionResults}. This underlying structure can be further exploited in the MPS encoding of the Fibonacci word.

\subsection{Encoding DFAO in Matrix Product States}\label{Sec: Sub 1.2}

Given a generalized automatic sequence, the aim is to construct its MPS representation $\hat W$ such that each sequence element corresponds directly to a tensor element
\begin{equation}
    W^{\sigma_1\sigma_2\cdots\sigma_L} = \sum_{\alpha_1,\dots,\alpha_{L-1}}[M_1]_{1\alpha_1}^{\sigma_1}[M_2]_{\alpha_1\alpha_2}^{\sigma_2}\cdots[M_L]^{\sigma_L}_{\alpha_{L-1}1},
    \label{Eq: MPS encoding}
\end{equation}
where each $M_l$ is a three-legged tensor having one physical index $\sigma_l$ and two bond indices $\alpha_{l-1}$ and $\alpha_l$. The dimensions of the bond indices are called the bond dimensions of $\hat W$. The maximum bond dimension, denoted by $\chi$, depends on the rule that connects sequence elements to the set of indices $\{\sigma_1, \sigma_2,\dots, \sigma_L\}$. This rule is referred to as \textit{MPS element encoding}.

If a sequence is $k$-automatic, a natural way to perform its MPS element encoding is to associate the $n$th sequence element to the tensor element $W^{\sigma_1\sigma_2\cdots\sigma_L}$ through the base-$k$ representation $n=(\sigma_1\sigma_2\cdots\sigma_L)_k$. The local tensors of $\hat W$ act as an internal logic that transforms the input digits into the desired sequence element mimicking the behavior of the generating DFAO with input alphabet $\Sigma_k$. The connection between automaton structure and MPS representation has been previously discussed in \cite{Crosswhite2008FiniteAlgorithms, Critch2013AlgebraicModels,Critch2014AlgebraicStates,Rabusseau2014LearningDecompositions,Li2024ConnectingLearning}. We use this correspondence directly by building $\hat W$ from the transition structure of the automaton generating the target sequence. Furthermore, the sequence does not have to be $k$-automatic in order to exploit this approach: it is enough that the sequence can be considered automatic in some numeration system.

The general approach for the construction is as follows. The numeration system in which the sequence is automatic defines the MPS element encoding. The dimensions of the physical and bond indices in \cref{Eq: MPS encoding} are determined by the generating DFAO: physical and bond dimensions equal the sizes of the input alphabet $\Sigma$ and set of states $S$, respectively \cite{Li2024ConnectingLearning}. As the number of states of a DFAO is independent of the input size, the maximum bond dimension of its MPS representation $\hat W$ is independent of the number of local tensors $L$. 

The DFAO generates the output by processing the input string $(\sigma_1\sigma_2\cdots\sigma_L)$ sequentially from the most significant digit $\sigma_1$ to the least significant digit $\sigma_L$. This process is further encoded into the local tensors of $\hat W$. The contraction in \cref{Eq: MPS encoding} can be performed from left to right as a sequence of vector-matrix products. Using a DFAO-based MPS encoding, at any point of the contraction the leftmost vector is a one-hot vector that describes the state of the process. Each of the elements of the state vector represents one possible state of the DFAO. The local bulk tensors serve as transition matrices that update the process state after reading the digit $\sigma_i$. To produce the final output, the last local tensor maps the final state into the corresponding output, reproducing the behavior of the output map $\tau$ of the DFAO.

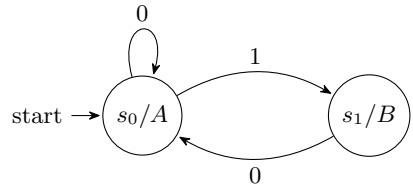
\begin{figure}[ht]
    \centering
    \begin{tikzpicture}[
        shorten >=1pt, 
        node distance=3cm, 
        on grid, 
        auto, 
        >={Stealth[round]} 
    ] 
        \node[state, initial] (s0) {$s_0/A$}; 
        \node[state] (s1) [right=of s0] {$s_1/B$}; 

        \path[->] 
        (s0) edge [loop above] node {0} (s0)
             edge [bend left]  node {1} (s1)
        (s1) edge [bend left]  node {0} (s0);
             
    \end{tikzpicture}
    \caption{The transition diagram of the DFAO generating the Fibonacci word. Each node is labeled as $s_i/\omega_i$, where $\omega_i\in\{A,B\}$ is the output corresponding to the final state being $s_i\in S$. The edges represent the state transitions labeled by the input bit. There is no transition from state $s_1$ to $s_1$ since consecutive ones are not allowed in the Fibonacci representation.}
    \label{Fig: fibonacci_dfao}
\end{figure}

To demonstrate the MPS encoding, we consider the Fibonacci word as an example. The $n$th symbol of the Fibonacci word, indexing from zero, can be recovered from the following rule:
\begin{enumerate}[nolistsep]
    \item Determine the last (least significant) digit of the Fibonacci representation of $n$.
    \item If the last digit is 0, then the $n$th symbol is $A$ and otherwise it is $B$.
\end{enumerate}
The Fibonacci word is generated by a 2-state DFAO that processes the digits of an input Fibonacci representation sequentially from the most to least significant digit and keeps track of the last digit processed so far. The states of the automaton $s_0$ and $s_1$ correspond to the last processed digit being 0 and 1, respectively. The transitions between the states are depicted in \cref{Fig: fibonacci_dfao}.

The initial state of the DFAO is $s_0$. After reading the first input digit $\sigma_1$ the automaton either remains in state $s_0$ or transitions to state $s_1$. The first local tensor is given by
\begin{equation*}
    [M_1]^0=
    \begin{bmatrix}
        1 & 0
    \end{bmatrix},
    \quad
    [M_1]^1=
    \begin{bmatrix}
        0 & 1
    \end{bmatrix},
\end{equation*}
where the superscript of the local tensor $M_1$ denotes the value of the first digit $\sigma_1$. During the tensor contraction, the current state of the automaton is represented by a one-hot vector whose $i$th element is equal to 1 if and only if the automaton is in state $s_i$. 

Since the DFAO is deterministic, each state has exactly one outgoing transition for every possible input digit. The local bulk tensors encode these transition rules and are given by
\begin{equation*}
    [M_l]^0=
    \begin{bmatrix}
        1 & 0 \\
        1 & 0 \\
    \end{bmatrix},
    \quad
    [M_l]^1=
    \begin{bmatrix}
        0 & 1\\
        0 & 0\\
    \end{bmatrix},
\end{equation*}
where $1<l<L$. The state transition from state $s_1$ to itself is forbidden
because valid Fibonacci representations do not contain consecutive ones.

The final local tensor depends on the last input digit $\sigma_L$ and the corresponding output.
It is given by
\begin{equation*}
    [M_L]^0=
    \begin{bmatrix}
        A & A
    \end{bmatrix}^\intercal
    ,\quad
    [M_L]^1=
    \begin{bmatrix}
        B & 0
    \end{bmatrix}^\intercal,
\end{equation*}
where $A$ and $B$ denote the two possible output values. With these local tensors, the $n$th sequence element, whose Fibonacci representation is $(\sigma_1\sigma_2\cdots\sigma_L)$,  corresponds directly to the tensor element $W^{\sigma_1\sigma_2\cdots\sigma_L}$. 

Although every sequence element corresponds to an element of $\hat W$, the converse is not true. The tensor $W$ accepts every binary string as an input, regardless of whether it is a valid Fibonacci representation. Tensor elements corresponding to invalid input strings are therefore assigned the value zero. 

The main drawback of the Fibonacci encoding compared with the binary encoding is the presence of invalid tensor elements. While $L$ binary digits can represent $2^L$ integers, the corresponding number of integers for Fibonacci encoding is given by the $(L+2)$nd Fibonacci number $F_{L+2}$. Consequently, an MPS representation based on the Fibonacci encoding must be longer than its binary counterpart to represent the same number of sequence elements. While the binary number of states grows asymptotically as $2^L$, the number of allowed Fibonacci states grows as $\varphi^L$.

This additional overhead is compensated by a significantly lower maximum bond dimension $\chi$. Since the Fibonacci word is not 2-automatic, its $n$th element cannot be determined directly from the binary expansion of $n$. As a result, an MPS with the binary encoding would require $\chi$ to grow with the system size, making numerical simulations presented in \cref{Sec: Sub 2.2} computationally infeasible. In contrast, the Fibonacci encoding yields an MPS representation with a constant maximum bond dimension, $\chi=2$, independently of the length of the encoded prefix.

\subsection{Generalizations} \label{Sec: Generalizations}
The Fibonacci quasicrystal provides the simplest paradigmatic example. 
This construction admits several natural generalizations. The first is the family of metallic-mean quasicrystals \cite{Thiem2011GeneralizedSystems}, also constructed from Sturmian words \cite{Lothaire2002SturmianWords,deSpinadel1999TheSpectra}, whose defining irrationals are the positive roots $\delta_m=(m+\sqrt{m^2+4})/2$ of $x^2-mx-1=0$, with $m\in\mathbb{N}$. Their continued-fraction expansions are $\delta_m=[m;m,m,m,\ldots]$, while the associated integer sequences obey the generalized Fibonacci recurrence $F_{n+2}^{(m)}=mF_{n+1}^{(m)}+F_n^{(m)}$. The corresponding substitution may be written as $A\to A^mB$ and $B\to A$, with inflation factor $\delta_m$. The cases $m=1$ and $m=2$ recover the Fibonacci and silver-mean, or Pell \cite{Baranwal2019CriticalSystem}, structures, respectively.
The numeration systems associated with these sequences generalize the Zeckendorf representation. Integers are expanded in the basis formed by the generalized Fibonacci numbers $F_n^{(m)}$, with digits and local admissibility constraints determined by the recurrence relation. Such representations can be processed by finite automata and therefore encoded exactly as MPSs of finite bond dimension. When the numeration digits take values in $\{0,\ldots,m\}$, the corresponding MPS representation has local physical dimension $m+1$.

More generally, metallic means constitute only a period-one subclass of quadratic irrationals \cite{deSpinadel1999TheSpectra}. For an arbitrary irrational slope $\alpha=[0;a_1,a_2,a_3,\ldots]$, the denominators $q_n$ of its continued-fraction convergents satisfy $q_{n+1}=a_{n+1}q_n+q_{n-1}$. They define the associated Ostrowski numeration system \cite{Ostrowski1922BemerkungenApproximationen,Berthe2001AutourDOstrowski,Hieronymi2018OstrowskiAutomata,Frougny2010NumberAutomata}, in which every non-negative integer admits a unique expansion $N=\sum_{n}d_nq_n$, subject to digit bounds and local carry constraints determined by the coefficients $a_n$. Quadratic irrational slopes have eventually periodic continued fractions, so their Ostrowski digit rules are eventually periodic and can be recognized by a finite automaton \cite{Frougny2010NumberAutomata}. Consequently, Sturmian sequences with quadratic slopes admit generalized automatic representations and hence exact finite-bond-dimension MPS encodings. Metallic-mean numeration, including the Fibonacci and silver-mean cases, is recovered when all continued-fraction coefficients are equal.

Another natural class of generalizations is provided by the $k$-bonacci words \cite{Carlitz1972FibonacciOrder,Kocabova2007AmbiguitySystem}. These are non-Sturmian, $k$-letter generalizations of the Fibonacci dynamical system, defined as the fixed points of the substitutions $0\to01$, $1\to02$, $\ldots$, $k-2\to0(k-1)$, and $k-1\to0$. Their inflation factor $\beta_k$ is the unique positive root larger than one of the polynomial $x^k-x^{k-1}-\cdots-x-1=0$. In the cut-and-project picture, the corresponding quasicrystal is obtained by projecting a $k$-dimensional lattice onto a one-dimensional physical subspace, with a $(k-1)$-dimensional internal space whose acceptance window is a Rauzy fractal \cite{Rauzy1982NombresSubstitutions,Berthe2010SubstitutionsTilings}. The orientation of the physical line, or equivalently of the complementary internal hyperplane, is determined by the Perron--Frobenius eigenvector $(1,\beta_k^{-1},\ldots,\beta_k^{-(k-1)})$ of the substitution matrix. Unlike the quadratic metallic-mean case, $\beta_k$ has algebraic degree $k$ and is therefore not characterized by an eventually periodic ordinary continued fraction. The corresponding generalization is instead formulated through multidimensional continued-fraction algorithms or, directly, through the $k$-dimensional substitution matrix \cite{Berthe2010SubstitutionsTilings}. The associated numeration basis is formed by the shifted $k$-bonacci numbers $U_l^{(k)}$, defined by $U_l^{(k)}=2^l$ for $0\leq l<k$ and $U_{l+k}^{(k)}=U_{l+k-1}^{(k)}+\cdots+U_l^{(k)}$, with $\beta_k=\lim_{l\to\infty}U_{l+1}^{(k)}/U_l^{(k)}$. Every non-negative integer admits a canonical binary expansion $N=\sum_l \sigma_lU_l^{(k)}$, with $\sigma_l\in\{0,1\}$ and no occurrence of $k$ consecutive ones \cite{Carlitz1972FibonacciOrder,Kocabova2007AmbiguitySystem}.

The $k$-bonacci word is therefore automatic in this numeration system and admits an exact MPS representation with local physical dimension $2$. A DFAO may be constructed using $k$ states that record the length of the terminal string of ones in the representation, leading to an MPS bond dimension $k$ and a $k$-symbol output alphabet. The cases $k=2$ and $k=3$ recover the Fibonacci and Tribonacci words, respectively.

In App.~\ref{App: other QC}, we present explicit MPS constructions for representatives of the two classes of generalizations discussed above, namely the silver-mean and Tribonacci words.

\section{Encoding a Quasicrystal in Hamiltonian Matrix Product Operators}\label{Sec: QC Ham MPO}
\begin{figure}[!hbt]
    \centering
    \includegraphics[]{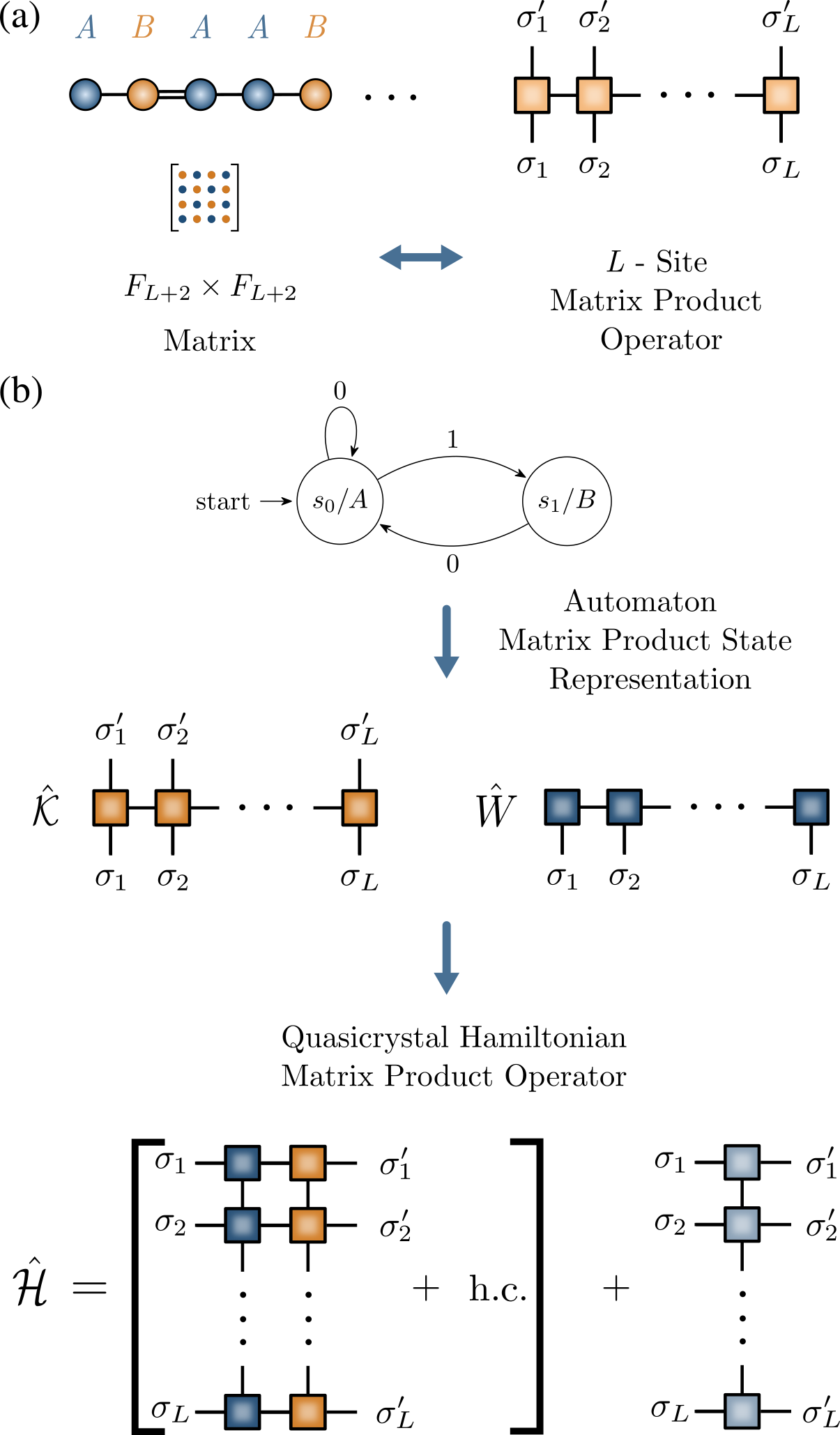}
    \caption{MPO representation of the Fibonacci chain. (a) An $F^{}_{L+2}\times F^{}_{L+2}$ Fibonacci chain Hamiltonian matrix is mapped to an $L$-site MPO by labeling sites with Fibonacci (Zeckendorf) bit strings. (b) The automaton generating the Fibonacci word defines basis for the shift MPO $\hat{\mathcal{K}}$ (the Zeckendorf basis) and the Fibonacci word MPS $\hat{W}$; their product $\hat{\mathcal{W}}\hat{\mathcal{K}}+ \mathrm{h.c.}$ gives a hopping term modulated by the Fibonacci sequence, where $\hat{\mathcal{W}}$ is the diagonal MPO representation of the Fibonacci MPS $\hat{W}$. Here, the potential is another arbitrary diagonal MPO. This methodology extends beyond the Fibonacci example shown here.}
    \label{fig: overview of method}
\end{figure}

We consider a generic nearest-neighbor tight-binding Hamiltonian of the form
\begin{equation}
    H = \sum_{n=0}^{N-1}v_n\ket{n}\bra{n}+\sum_{n=0}^{N-2}t_n(\ket{n}\bra{n+1}+\ket{n+1}\bra{n}),
    \label{Eq: matrix_Ham}
\end{equation}
where $t_n$ denotes the hopping amplitudes between adjacent sites $n$ and $n+1$ and $v_n$ is the on-site potential at site $n$. Quasiperiodicity is introduced to the model by modulating either the hopping amplitudes $t_n$ or the on-site potentials $v_n$ according to an aperiodic sequence. As the system size $N$ increases, the operator becomes impossible to store explicitly. Our approach is to represent the Hamiltonian in \cref{Eq: matrix_Ham} as an MPO by exploiting the MPS representation of the underlying aperiodic sequence. 

We consider both diagonal and off-diagonal quasiperiodic models. A model is called diagonal (off-diagonal) when the on-site potentials (hopping amplitudes) are modulated according to an aperiodic sequence, while the hopping-amplitudes (on-site potentials) remain constant. The corresponding Hamiltonian can be written as $H=V+TK+(TK)^\dagger$, where $V$ and $T$ are diagonal matrices containing the on-site potentials and hopping amplitudes, respectively, and $K$ is the upper shift-matrix. Following the approach of \cite{Sun2025Self-consistentSites,Fumega2025CorrelatedAlgorithm,Moustaj2026TensorSystems,Antao2025TensorMosaics,Moustaj2026Tensor-networkSites,Sun2026Real-spaceNetworks,Antao2026TensorApplications}, the MPO representation $\hat{\mathcal{H}}$ is constructed by adopting tensor network representations $\hat{\mathcal{V}}$, $\hat{\mathcal{T}}$ and $\hat{\mathcal{K}}$ for the matrices $V$, $T$ and $K$, respectively. A schematic of this methodology is shown in \cref{fig: overview of method}, where $\hat{\mathcal{T}}=\hat{\mathcal{W}}$.

The central idea of our methodology is to employ a sequence-dependent MPS element encoding. The encoding is determined by the numeration system in which the modulation sequence defining the Hamiltonian is (generalized) automatic. Accordingly, metallic mean words are encoded using Ostrowski numeration systems, whereas $k$-bonacci words are encoded using $k$-bonacci number based numeration systems. 

Each lattice index $n$ is associated with its sequence-dependent representation $(\sigma_1\sigma_2\cdots\sigma_L)_Q$, where $\sigma_i\in\{0,\dots,d-1\}$ for all $1\leq i\leq L$, and $Q$ is used to denote a general representation. For $k$-bonacci words, $d=2$, whereas for a metallic-mean word with parameter $m$, $d=m+1$. The number of representable lattice sites depends on the number of digits $L$, the local dimension $d$, and the constraints of the chosen numeration system.

Using this encoding, the Hamiltonian is then represented as an MPO that acts on the full $d^L$-dimensional tensor product space spanned by the states $\ket{\sigma_1\sigma_2\cdots\sigma_L}$:
\begin{equation}
\hat{\mathcal{H}}=\sum_{\{\boldsymbol{\sigma},\boldsymbol{\sigma}'\}}\Gamma_1^{\sigma_1\sigma_1'}\Gamma_2^{\sigma_2\sigma_2'}\cdots\Gamma_L^{\sigma_L\sigma_L'}\ket{\sigma_1\sigma_2\cdots\sigma_L}\bra{\sigma'_1\sigma'_2\cdots\sigma'_L},
    \label{Eq: H_MPO}
\end{equation}
where each $\Gamma_i$ is a four-legged tensor. The matrix element $H_{n,m}$ is mapped to a tensor element $\mathcal{H}^{\sigma, \sigma'}$, with $n=(\sigma_1\sigma_2\cdots\sigma_L)_Q$ and $m=(\sigma_1'\sigma_2'\cdots\sigma_L')_Q$. Although the Hamiltonian $\hat{\mathcal{H}}$ is defined on the full $d^L$-dimensional Hilbert space, only the basis states corresponding to valid strings in the chosen numeration system represent physical lattice sites. Therefore, 
a tensor element $\mathcal{H}^{\sigma, \sigma'}$ for which either $\sigma$ or $\sigma'$ is an invalid string should not affect the computation of any physical observable. To this end, depending on the intended application, it may be advantageous to project the Hamiltonian onto a subspace spanned by the valid states after its construction. Another option is to develop specific algorithms that disregard physically meaningless states when extracting the final result, as is done by the trace operation in \cref{Sec: Sub 2.2}.

The constructions of $\hat{\mathcal{V}}$ and $\hat{\mathcal{T}}$ follow in a similar manner. As discussed in \cref{Sec: Generalizations}, both $k$-bonacci and metallic-mean words admit MPS representations with finite bond dimensions. By introducing an auxiliary physical index $\sigma_i'$ to each MPS tensor $M_i^{\sigma_i}$, we obtain a four-legged tensor $N_i^{\sigma_i\sigma_i'}$. The diagonal structure is then enforced by contracting each tensor with the Kronecker delta $\delta_{\sigma_i \sigma_i'}$. These resulting tensors define the MPO representation of either $\hat{\mathcal{V}}$ or $\hat{\mathcal{T}}$, depending on whether the diagonal or off-diagonal quasiperiodic model is considered, as shown in \cref{fig: overview of method}. 

To obtain the form in \cref{Eq: H_MPO}, it remains to construct $\hat{\mathcal{K}}$. The operator $\hat{\mathcal{K}}$ implements the map $\ket{n}\mapsto\ket{n-1}$ within the sequence-dependent representation, thereby mimicking the behavior of its matrix counterpart $K$. The general approach for the construction of $\hat{\mathcal{K}}$ is as follows. The first step is to determine the digit-wise transformation relating the representations of $n$ and $n-1$ in the same numeration system. For example, in the binary system, the least significant digit equal to 1 of the binary expansions of $n$ is changed to 0, while the less significant trailing zeros are changed to 1. Then the digit-wise subtraction is translated into the tensor formalism by modifying the local tensors of the product state $\ket{n}=\ket{\sigma_1}\otimes\ket{\sigma_2}\otimes\cdots\otimes\ket{\sigma_L}$, where each local tensor represents a single digit. To this end, each possible digit transformation is associated with a corresponding local operator. Finally, $\hat{\mathcal{K}}$ is constructed as an operator sum that encodes the subtraction procedure at the digit-level, that is, by acting on local tensors of $\ket{n}$.

The general quasiperiodic Hamiltonian $\hat{\mathcal{H}}$ is obtained as
\begin{equation}
    \hat{\mathcal{H}}=\hat{\mathcal{V}}+\hat{\mathcal{T}}\hat{\mathcal{K}}+(\hat{\mathcal{T}}\hat{\mathcal{K}})^\dagger,
    \label{Eq: H_mpo_sum}
\end{equation}
where the element encoding and the shift tensor $\hat{\mathcal{K}}$ both depend on the modulating sequence defining the Hamiltonian. To illustrate the construction, we consider the encoding of a Fibonacci quasicrystal in more detail, while the details for the silver-mean \cite{Cerovski2005SpectralDimensions} and Tribonacci \cite{Krebbekx2023MultifractalChains} quasicrystals can be found in App.~\ref{App: other QC}.

\subsection{The Fibonacci chain} \label{Sec: Sub 2.1}

The Fibonacci Hamiltonian $\hat{\mathcal{H}}_F$ has the same form as the generic quasiperiodic Hamiltonian in \cref{Eq: H_mpo_sum}. The crucial step is to use the Fibonacci numeration system in which the Fibonacci word is automatic. That numeration system determines both the element encoding and the subtraction rule implemented into the shift tensor $\hat{\mathcal{K}}$.

The Hamiltonian  $\hat{\mathcal{H}}_F$ is encoded with the Fibonacci element encoding. Each matrix element $[H_{F}]_{n,m}$ is mapped to a tensor element $\mathcal{H}_F^{\sigma, \sigma'}$ with $n=(\sigma_1\sigma_2\cdots\sigma_L)_F$ and $m=(\sigma_1'\sigma_2'\cdots\sigma_L')_F$. Although the operator $\hat{\mathcal{H}}_F$ acts on the full $2^L$-dimensional Hilbert space, only the basis states corresponding to valid Fibonacci strings represent physical lattice sites. Therefore, a Hamiltonian with $L$ tensor cores represents a system with $F_{L+2}$ physical sites.

To construct $\hat{\mathcal{H}}_F$, it remains to construct $\hat{\mathcal{K}}$. The operator $\hat{\mathcal{K}}$ implements the map $\ket{n}\mapsto\ket{n-1}$, where the state $\ket{n}$ is labeled by the integer $n=(\sigma_1\sigma_2\cdots\sigma_L)_F$ written in the Fibonacci representation. It is constructed using local operators, each acting on a state representing specific Fibonacci digit $\sigma_i\in\{0,1\}$, where the lower index $i$ denotes the digit position. The local operators used in the construction are the ladder operators $\hat\tau_i^+=\ket{1}\bra{0}_i$ and $\hat\tau_i^-=\ket{0}\bra{1}_i$, together with the projection operator $\hat p_i^0=\ket{0}\bra{0}_i$.

The operator $\hat{\mathcal{K}}$ can be expressed concisely as 
\begin{equation}
    \hat{\mathcal{K}}=\sum_{i=1}^L\hat\tau_i^-\bigotimes_{j>i}O_j,
    \quad O_j=
    \begin{cases}
    \hat\tau_j^+ \quad \textrm{if} \quad (j-i) \bmod 2 = 1\\
    \hat p_j^0 \quad \mathrm{otherwise}
    \end{cases},
    \label{eq: K opsum}
\end{equation}
where the identity operators acting on the remaining sites are implicit. The $i$th term of the operator sum in \cref{eq: K opsum} performs the subtraction when the least significant digit 1 in the Fibonacci representation of the input state label $n$ is at position $i$. The construction of the $i$th operator follows a procedure similar to that used for the silver-mean and Tribonacci models in App.~\ref{App: silver-mean ham} and App.~\ref{App: triboancci Ham}, respectively.

Operator $\hat{\mathcal{K}}$ is used to build a tensor network Hamiltonian with open boundary conditions. However, periodic boundary conditions can be introduced by adding to the right hand side of \cref{eq: K opsum} an additional operator
\begin{equation*}
    \hat{\tau}_1^+\otimes \hat{p}_2^0\otimes\hat{\tau}_3^+\otimes \hat{p}_4^0\otimes\hat{\tau}_5^+\otimes \hat{p}_6^0\otimes\cdots, 
\end{equation*}
where the local operators alternate between $\hat{\tau}^+$ and $\hat{p}^0$. The given operator implements the map $\ket{0}\mapsto\ket{N-1}$ without further changing the behavior of $\hat{\mathcal{K}}$. The other boundary is taken into account in the Hermitian conjugate of $\hat{\mathcal{K}}$.

Unlike the MPS representation of the Fibonacci word introduced in \cref{Sec: Sub 1.2}, the MPO representation $\hat{\mathcal{H}}_F$ does not necessarily assign zero to invalid tensor elements. This is because the shift tensor $\hat{\mathcal{K}}$ does not, in general, annihilate invalid states on which it acts. Depending on the intended application, it may therefore be advantageous to project the Hamiltonian onto the Fibonacci subspace after construction. The corresponding projection tensor is obtained by first constructing the Fibonacci word MPS  using the convention $A=B=1$ and then promoting it to a diagonal MPO. The maximum bond dimension of the resulting Hamiltonian is approximately 10, depending on the model, parameter values, system size and whether the projector is applied. 

\subsection{Spectral density evaluation}\label{Sec: Sub 2.2}

\begin{figure*}[t]
    \centering
    \includegraphics[]{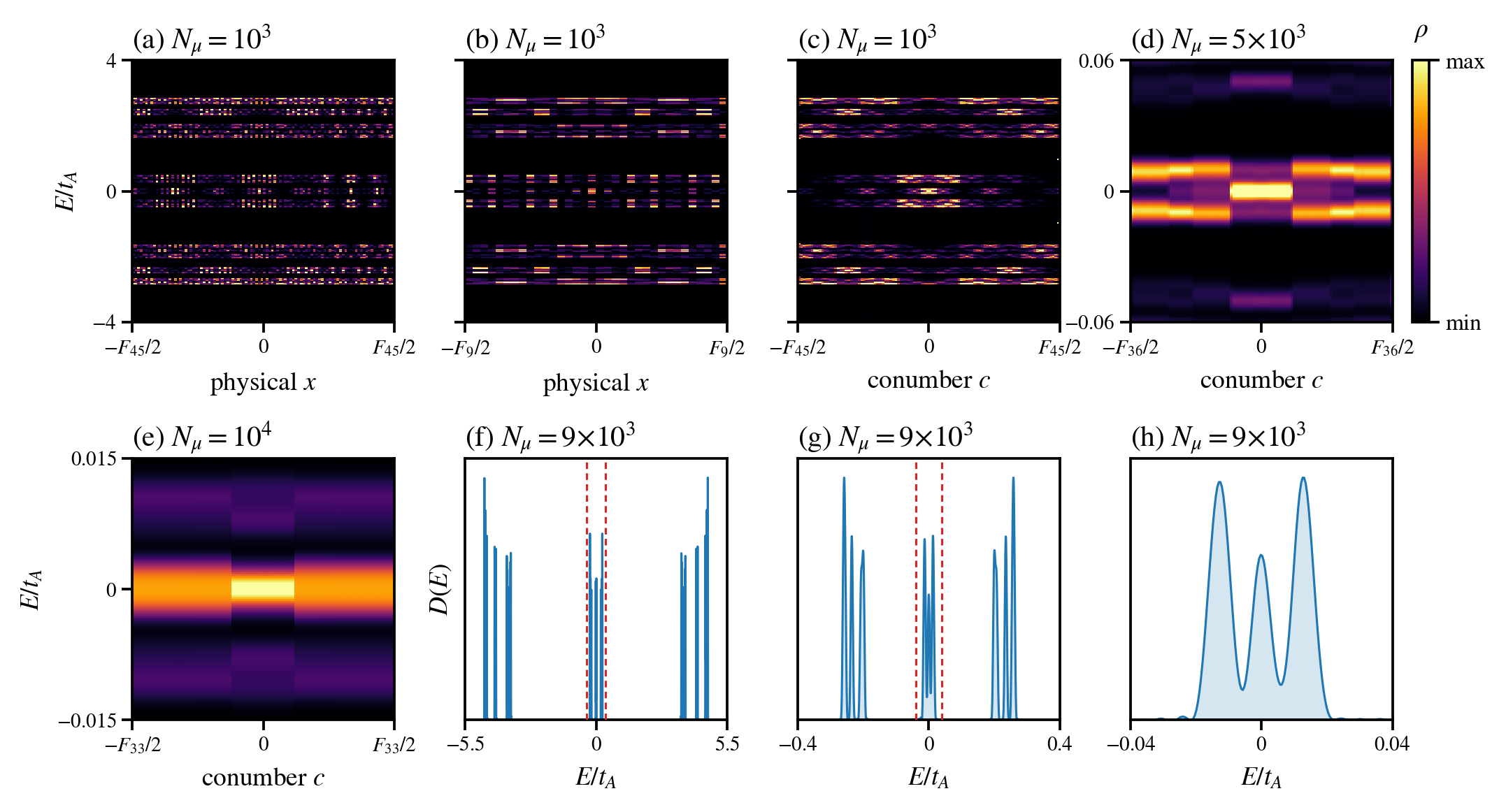}
    \caption{(a) LDOS of a Fibonacci chain with $t_A=1$ and $t_B=2$, containing  $F_{45}>10^9$ atoms. (b) Enlargement of the central region, showing $F_9=34$ atoms. (c) LDOS with the sites rearranged according to their local environments. (d,e) Enlargements of the central regions corresponding to the 3rd and 4th successive zooms, respectively. (f) DOS of a Fibonacci chain with $t_A=1$ and $t_B=4$, containing $F_{27}>10^5$ atoms. (g,h) Enlarged windows of the energy ranges indicated by the red lines in panels (f) and (g), respectively.}
    \label{fig: Fibonacci DOS+LDOS}
\end{figure*}

The Kernel Polynomial Method (KPM) \cite{Weie2006TheMethod} can be used to compute spectral quantities of a quasiperiodic Hamiltonian $\hat{\mathcal{H}}$ within the tensor network formalism. The central idea is to approximate the Dirac-delta operator,
\begin{equation}
    \delta(E-\hat{\mathcal{H}})= \frac{1}{\pi\sqrt{1-E^2}}\left[ \mathbb{1}+2\sum_{n=1}^\infty \hat{\mu}_nT_n(E) \right],
    \label{eq: Dirac-delta}
\end{equation}
where $\hat{\mathcal{H}}$ is the rescaled Hamiltonian, whose spectrum lies within the open interval $(-1,1)$. Here, $T_n(E)$ is the $n$th Chebyshev polynomial defined recursively as $T_n(E)=2ET_{n-1}(E)-T_{n-2}(E)$ with the initial conditions $T_0(E)=1$ and $T_1(E)=E$. The $n$th Chebyshev moment operator $\hat{\mu}_n=T_n(\hat{\mathcal{H}})$ is generated using the operator analog of the same recurrence relation. 

In practice, the expansion in \cref{eq: Dirac-delta} is approximated by truncating the Chebyshev expansion to $N_\mu$ terms. Truncating the infinite sum introduces Gibbs oscillations, which can be damped using a smoothing kernel. We use the Jackson kernel for this purpose, which provides convergence of order $1/N_{\mu}$ \cite{Jackson1912OnPolynomials}.

The computation of the Dirac-delta operator follows the same procedure as in \cite{Antao2026TensorApplications, Fumega2025CorrelatedAlgorithm, Sun2025Self-consistentSites,Antao2025TensorMosaics,Moustaj2026Tensor-networkSites,Moustaj2026TensorSystems,Sun2026Real-spaceNetworks}, where the KPM is applied to a binary-encoded tensor network Hamiltonian. The quasiperiodic encoding only enters when the density of states (DOS) or local density of states (LDOS) is extracted from the operator. The DOS is obtained as $D(E)=\textrm{Tr}_Q[\delta(E-\hat{\mathcal{H}})]$, where the trace is taken only over states corresponding to physical lattice sites, rather than over the full Hilbert space. A recursive algorithm for computing the restricted trace is described in detail in App.~\ref{App: restricted trace}.  The LDOS at site $n$ is computed as $\rho(E, n)=\bra{n}\delta(E-\hat{\mathcal{H}})\ket{n}$, where $\ket{n}$ denotes the MPS representing physical lattice site $n$ in the Fibonacci element encoding.

We leverage the intrinsic compressibility of tensor-network representations throughout the numerical procedure. All tensor-network operations are implemented using the ITensors library \cite{Fishman2022CodebaseITensor,Fishman2022TheCalculations}, with intermediate MPS and MPO objects compressed after contractions and operator applications. The resulting approximation scheme is controlled by the maximum bond dimension $\chi_{\rm max}$ and the truncation tolerance $\mathrm{tol}$, which respectively limit the complexity of the tensor-network representation and discard singular values below a prescribed threshold.

We now use the tensorized KPM methodology to compute the DOS and LDOS of Fibonacci chains. We consider the off-diagonal Fibonacci model, where the hopping amplitudes follow the Fibonacci word with parameters $t_A$ and $t_B$, while the on-site potentials are set to zero. The LDOS and total DOS are computed with different model parameters, Chebyshev recursion schemes and truncation parameters, as detailed below. 

For the DOS, we consider a Fibonacci chain of $N=F_{27}= 196,418$ atoms with open boundary conditions and hopping parameters $t_A=1$ and $t_B=4$. To avoid storing all the moment operators $\hat\mu_n$, we compute their Fibonacci restricted traces during the Chebyshev recursion and only store three consecutive Chebyshev polynomials at once.  During the recursion, the maximum bond dimension of the moment operators is set to $\chi_{\rm max}=800$ and the tolerance is set to $\rm tol = 10^{-10}$.

The energy spectrum of the off-diagonal Fibonacci model is symmetric and has three main clusters, each of which recursively consists of three sub-clusters \cite{Jagannathan2021TheMultifractality}.  This fractal nature of the spectrum can be resolved with KPM, as depicted in \cref{fig: Fibonacci DOS+LDOS} (f)-(h).

For the LDOS, we instead consider a Fibonacci chain of $N=F_{45}=1,134,903,170$ atoms with periodic boundary conditions and hopping parameters $t_A=1$ and $t_B=2$. Rather than propagating an MPO through the Chebyshev recursion, we choose a reference state $\ket{n}$ and propagate the MPS $T_n(\hat{\mathcal{H}})\ket{n}$. This approach corresponds to the MPS pathway of the KPM described in \cite{Antao2026TensorApplications,Moustaj2026Tensor-networkSites}. Compared to the MPO propagation, the maximum bond dimension grows more slowly, at the cost of requiring several sweeps. During the recursion, the maximum bond dimension is set to $\chi_{\rm max}= 400$  and tolerance is set to $\rm tol = 10^{-8}$. However, the bond dimension in all cases does not grow beyond $\chi=100$.

The hierarchical structure of the LDOS can be revealed by rearranging the sites according to their local environments using a conumbering scheme \cite{Sire1990ExcitationQuasicrystals}. Throughout the LDOS computations, we employ the following reordering. The raw conumbers are computed as $c'_i= iF_{L} \bmod F_{L+2}$, where $0\leq i <F_{L+2}$. We then apply a cyclic shift to the raw conumbers to move the modular cut so that the atomic (AA) sites form the central block.

The LDOS is depicted in the spatial ordering of the lattice sites in \cref{fig: Fibonacci DOS+LDOS} (a)-(b). Reordering the sites according to their local environments reveals a self-similar structure of the LDOS: the pattern of the whole chain of length $F_N$ is reproduced in the central window of length $F_{N-3n}$, where $n$ is an integer. This self-similarity is illustrated in \cref{fig: Fibonacci DOS+LDOS} (c)-(e), where panels (d) and (e) reproduce the pattern of panel (c) after 3 and 4 successive zooms, respectively. The spatial resolution of the LDOS is limited by the number of moments $N_\mu$. Consequently, resolving the repeated pattern requires increasing $N_\mu$ with the number of magnifications, which is evident in the panels (c)-(e), where $N_\mu$ is increased from $1,000$ to $10,000$.

\section{Conclusion} \label{Sec: Conclusion}

The construction of quasicrystals rests on a discrete set of substitution rules, which, when written in the appropriate numeration system, take the form of a DFAO that reads the digits of a site index and returns the corresponding letter of the quasiperiodic sequence. By leveraging the equivalence between such an automaton and an MPS, and combining this equivalence with
the construction of tight-binding Hamiltonians from binary encodings in the tensor-train language, we obtain a generalized construction principle for one-dimensional quasicrystals. We specifically develop a scheme for two families, the metallic-mean quasicrystals and the $k$-bonacci quasicrystals. In each case, the substitution rule fixes the numeration system, the numeration system fixes the automaton, the automaton fixes the MPS, and the tight-binding Hamiltonian follows as an exact MPO. The transition matrices of the DFAO are the MPS tensors, and the bond dimension is set by the number of automaton states rather than by the system size, allowing for a very efficient encoding of the Hamiltonian operator. We provide explicit constructions of the Fibonacci, silver-mean, and Tribonacci chains and use them to simulate systems with more than $10^9$ sites, extracting spectral densities using the kernel polynomial method. 
Reaching very large sizes with this method could enable more efficient evaluation of the defining features of aperiodic systems, such as critical states, multifractality, and the hierarchical organization of the spectrum, which emerge only asymptotically and are masked by finite-size effects at conventionally accessible sizes. As an outlook, the natural future step is to extend the same idea to 2D, where system sizes are more prohibitive to simulate. 
	
\textbf{Acknowledgments}
We acknowledge the computational resources provided by the Aalto Science-IT project
and the financial support from InstituteQ, 
the
Research Council of Finland (project No. 370912), 
the Finnish Ministry
of Education and Culture through the Quantum Doctoral Education Pilot Program (QDOC VN/3137/2024-OKM-4), the
Finnish Quantum Flagship (project No. 358877, Aalto University),
the Finnish Centre of Excellence in Quantum Materials QMAT (No. 374166),
and the ERC Consolidator Grant ULTRATWISTROICS (Grant agreement no.
101170477). 
We thank T. Ant\~ao and Y. Sun for useful discussions. 
The code used for this work can be consulted at \cite{millarepo}.

\onecolumngrid
\appendix
\newpage

\section{Recursive algorithm for a restricted trace} \label{App: restricted trace}

In \cref{Sec: Sub 2.2}, evaluating the DOS requires taking a trace only over states corresponding to physical lattice sites instead of over the full $d^L$-dimensional Hilbert space, where $d$ is the local dimension. The modeled quasiperiodic chain defines the numeration system, which is then used to encode the Hamiltonian as an MPO, $\hat{\mathcal{H}}$.  A general representation $(\sigma_1\sigma_2\cdots\sigma_L)_Q$ corresponds to a physical lattice site if it is a valid representation in that chain-dependent numeration system.

The restricted trace of an MPO $\hat{\mathcal{A}}$ with $L$ sites is defined as 
\begin{equation}
    \textrm{Tr}_Q(\hat{\mathcal{A}})=\sum_{n\in \textrm{Qr}_L}\bra{n}\hat{\mathcal{A}}\ket{n},
    \label{eq: restricted trace}
\end{equation}
where the summation is performed over the set of valid representations of length $L$ in the chain-dependent numeration system. A direct evaluation of the restricted trace using \cref{eq: restricted trace} requires computing as many inner products within the tensor network formalism as there are valid representations of $L$ digits. The number of such representations tends to grow exponentially with the number of tensor cores. However, the trace can instead be obtained directly from the local tensors of $\hat{\mathcal{A}}$ as
\begin{equation}
    \textrm{Tr}_Q(\hat{\mathcal{A}})=\sum_{n\in \textrm{Qr}_L}[A_1]^{\sigma_1\sigma_1}[A_2]^{\sigma_2\sigma_2}\cdots[A_L]^{\sigma_L\sigma_L},
    \label{eq: trace local}
\end{equation}
where $n=(\sigma_1\sigma_2\cdots\sigma_L)_Q$ and $\textrm{Qr}_L$ is the set of valid representations of length $L$. 

To incorporate the admissibility condition of the numeration system, the set $\textrm{Qr}_L$ is divided into subsets, denoted by $\textrm{Qr}^x_L$. The number of subsets equals the number of states required to encode the admissibility condition. The trace in \cref{eq: trace local} can then be written as
\begin{equation*}
    \textrm{Tr}_Q(\hat{\mathcal{A}})=\sum_x\sum_{n\in \textrm{Qr}^x_L}[A_1]^{\sigma_1\sigma_1}[A_2]^{\sigma_2\sigma_2}\cdots[A_L]^{\sigma_L\sigma_L}:=\sum_x R_L^x,
\end{equation*}
where each sub-result $R_L^x$ can be evaluated recursively such that the admissibility conditions of the numeration system are satisfied. This formulation allows the restricted trace to be computed directly from the local tensors of $\hat{\mathcal{A}}$ providing better computational efficiency than the direct formulation in \cref{eq: restricted trace}. The construction is illustrated below for the Fibonacci, silver-mean and Tribonacci traces.

\subsection{Fibonacci trace} \label{App: fibo trace}

The set of valid Fibonacci representations $\textrm{Fib}_L$ consists of binary strings that do not contain two consecutive ones. To incorporate this restriction, the set $\textrm{Fib}_L$ is divided into two subsets $\textrm{Fib}_L^0$ and $\textrm{Fib}_L^1$ containing strings ending in 0 and 1, respectively. The Fibonacci trace can then be written as 
\begin{equation*}
     \textrm{Tr}_F(\hat{\mathcal{A}})=\sum_{n\in \textrm{Fib}_{L}^0}[A_1]^{\sigma_1\sigma_1}[A_2]^{\sigma_2\sigma_2}\cdots[A_L]^{00}+\sum_{n\in \textrm{Fib}_L^1}[A_1]^{\sigma_1\sigma_1}[A_2]^{\sigma_2\sigma_2}\cdots[A_L]^{11} := R^0_{L}+R^1_L.
\end{equation*}
The sub-results satisfy the recurrence relations
\begin{align*}
    R^0_L&=(R^0_{L-1}+R^1_{L-1})[A_L]^{00}, \\
    R^1_L&=R^0_{L-1}[A_L]^{11}
\end{align*}
with the initial conditions $R^0_1=[A_1]^{00}$ and $R^1_1=[A_1]^{11}$. 

\subsection{Silver-mean trace} \label{App: sm trace}

The set of valid silver-mean representations $\textrm{Sm}_L$ consists of all ternary strings satisfying the admissibility condition $\sigma_l=2 \Rightarrow\sigma_{l+1}=0$ for all $1\leq l<L$. To incorporate this restriction, the set $\textrm{Sm}_L$ is divided into three subsets $\textrm{Sm}_L^0$,  $\textrm{Sm}_L^1$ and  $\textrm{Sm}_L^2$ containing strings ending in $0$, $1$ and $2$, respectively. 

Similarly, the trace can be divided into a sum of three terms
\begin{equation*}           
\textrm{Tr}_{S}(\hat{\mathcal{A}}):=R^0_{L}+R^1_L+R^2_L,
\end{equation*}
where each sub-result $R^x_L$ is obtained by evaluating the right-hand side of \cref{eq: trace local} over $\textrm{Sm}_L^{x}$ instead of $\textrm{Sm}_L$. Taking the silver-mean admissibility condition into account, the sub-results satisfy
\begin{align*}
    R^0_L &= (R^0_{L-1}+R^1_{L-1}+R^2_{L-1})[A_{L}]^{00}\\
    R^1_L &= (R^0_{L-1}+R^1_{L-1})[A_L]^{11}\\
    R^2_L &= (R^0_{L-1}+R^1_{L-1})[A_L]^{22}
\end{align*}
with the initial conditions $R^0_1=[A_1]^{00}$, $R^1_1=[A_1]^{11}$ and $R^2_1=[A_1]^{22}$.

\subsection{Tribonacci trace} \label{App: Tribo trace}

The set of valid Tribonacci representations $\textrm{Trib}_L$ consists of all binary strings that do not contain three consecutive ones. To incorporate this restriction, the set $\textrm{Trib}_L$ is divided into three subsets $\textrm{Trib}_L^0$,  $\textrm{Trib}_L^{01}$ and  $\textrm{Trib}_L^{11}$ containing strings ending in $00/10$, $01$ and $11$, respectively. Similarly, the trace can be divided into a sum of three terms
\begin{equation*}           
\textrm{Tr}_{T}(\hat{\mathcal{A}}):=R^0_{L}+R^{01}_L+R^{11}_L,
\end{equation*}
where each sub-result $R^x_L$ is obtained by evaluating the right-hand side of \cref{eq: trace local} over $\textrm{Trib}_L^{x}$ instead of  $\textrm{Trib}_L$. Taking the restriction of the Tribonacci numeration system into account, the sub-results satisfy
\begin{align*}
    R^0_L &= (R^0_{L-1}+R^{01}_{L-1}+R^{11}_{L-1})[A_{L}]^{00}\\
    R^{01}_L &= R^{0}_{L-1}[A_L]^{11}\\
    R^{11}_L &= R^{01}_{L-1}[A_L]^{11}
\end{align*}
with the initial conditions $R^0_1=[A_1]^{00}$, $R^{01}_1=[A_1]^{11}$ and $R^{11}_1=0$.

\section{Other examples of quasicrystals}\label{App: other QC}

We present two examples from the two classes of generalizations discussed in \cref{Sec: Generalizations}: the silver-mean quasicrystal and the Tribonacci quasicrystal. 

\subsection{The silver-mean quasicrystal}

\subsubsection{MPS representation}

As another example in the Sturmian class beyond the Fibonacci quasicrystal presented in the main text, we explicitly construct the MPS for the silver-mean quasicrystal corresponding to $m=2$, and which has a natural representation in the Ostrowski numeration system. The silver-mean word is generated by the substitution rule $A\to AAB$ and $B\to A$.
Its Sturmian slope is $\alpha_{\mathrm{sm}}=1/(2+\sqrt{2})=[0;3,\overline{2}]$, which is related to the silver mean through $\alpha_{\mathrm{sm}}=1/(\delta_2+1)$. The denominators of its continued-fraction convergents satisfy $q_0=1$, $q_1=3$, and $q_{l+1}=2q_l+q_{l-1}$, producing the sequence $1,3,7,17,41,\ldots$. These numbers are related to the Pell numbers $P_l$ by $q_l=P_l+P_{l+1}$ and form the natural Ostrowski numeration basis of the silver-mean word.

Every non-negative integer $n$ has a unique representation $n=\sum_{l=0}^{L-1}\sigma_{L-l}q_l$, where $\sigma_l\in\{0,1,2\}$ and the digits satisfy the admissibility condition $\sigma_l=2\Rightarrow\sigma_{l+1}=0$ for $0\leq l<L$. Equivalently, when the digits are written from the most to the least significant position, every digit $2$ must be followed by a $0$. The $n$th symbol of the silver-mean word, indexing from zero, can then be recovered from the following rule:
\begin{enumerate}[nolistsep]
\item Determine the last, or least significant, digit of the silver-mean representation of $n$.
\item If the last digit is $2$, then the $n$th symbol is $B$; otherwise, it is $A$.
\end{enumerate}

\begin{figure}[t]
    \centering
    \includegraphics[]{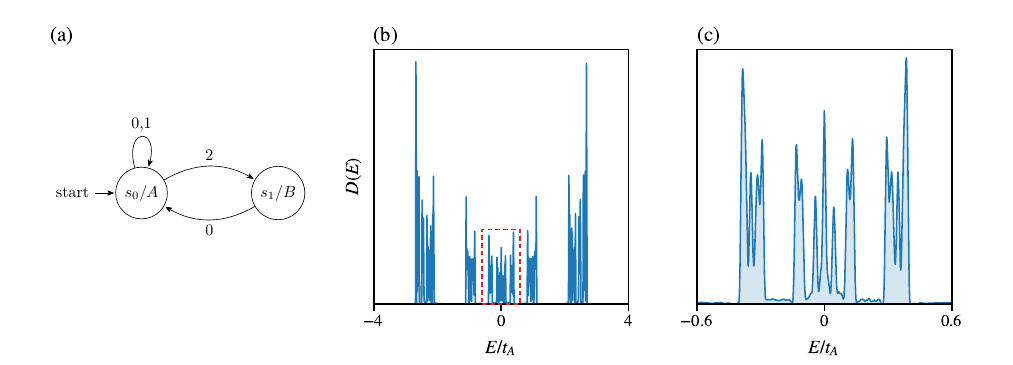}
    \caption{(a) Transition diagram of the DFAO generating the silver-mean word. Each node is labeled as $s_i/\omega_i$, where $\omega_i\in\{A,B\}$ is the output associated with the final state $s_i$. Transitions labeled by $1$ or $2$ from $s_1$ are absent because a digit $2$ must be followed by $0$ in a valid silver-mean representation. (b) DOS of the silver-mean chain 
    containing $N>10^7$ atoms computed using the KPM with $N_\mu=2,300$. (c) Enlarged view of the central region indicated by the red rectangle in panel (b).}
    \label{fig: sm automaton+DOS}

\end{figure}

The silver-mean word is generated by a two-state DFAO that processes the digits of the input representation sequentially from the most to the least significant digit. The state $s_0$ indicates that the last processed digit belongs to $\{0,1\}$, whereas $s_1$ indicates that the last processed digit is $2$. The corresponding outputs are $A$ and $B$, respectively. The transitions between the states are depicted in \cref{fig: sm automaton+DOS} (a).

The initial state of the DFAO is $s_0$. After reading the first input digit, the automaton remains in $s_0$ when $\sigma_1=0$ or $\sigma_1=1$, and transitions to $s_1$ when $\sigma_1=2$. The first local tensors are therefore
\begin{equation*}
[M_1]^0=
\begin{bmatrix}
1 & 0
\end{bmatrix},
\qquad
[M_1]^1=
\begin{bmatrix}
1 & 0
\end{bmatrix},
\qquad
[M_1]^2=
\begin{bmatrix}
0 & 1
\end{bmatrix}.
\end{equation*}
During the tensor contraction, the current state of the automaton is represented by a one-hot vector whose $i$th component is equal to $1$ if and only if the automaton is in state $s_i$.

The local bulk tensors encode the transition rules of the DFAO and are given by
\begin{equation*}
[M_l]^0=
\begin{bmatrix}
1 & 0 \\
1 & 0
\end{bmatrix},
\qquad
[M_l]^1=
\begin{bmatrix}
1 & 0 \\
0 & 0
\end{bmatrix},
\qquad
[M_l]^2=
\begin{bmatrix}
0 & 1 \\
0 & 0
\end{bmatrix},
\end{equation*}
where $1<l<L$. The vanishing second rows of $[M_l]^1$ and $[M_l]^2$ encode the fact that, after reading a digit $2$, the next digit must be $0$. The tensors therefore define the DFAO on the language of valid silver-mean representations; transitions corresponding to invalid digit strings are assigned zero weight.

The final local tensor depends on the last input digit $\sigma_L$ and maps the resulting automaton state to its corresponding output. It is given by
\begin{equation*}
[M_L]^0=
\begin{bmatrix}
A\
A
\end{bmatrix}^\intercal,
\qquad
[M_L]^1=
\begin{bmatrix}
A\
0
\end{bmatrix}^\intercal,
\qquad
[M_L]^2=
\begin{bmatrix}
B\
0
\end{bmatrix}^\intercal.
\end{equation*}
With these local tensors, the $n$th element of the silver-mean word, whose silver-mean representation is $(\sigma_1\sigma_2\cdots\sigma_L)$, is given directly by the tensor element $W^{\sigma_1\sigma_2\cdots\sigma_L}$. The resulting MPS has bond dimension $2$, equal to the number of DFAO states, and local physical dimension $3$, corresponding to the three possible numeration digits.

\subsubsection{Silver-mean Hamiltonian} \label{App: silver-mean ham}

The silver-mean Hamiltonian  $\hat{\mathcal{H}}_{S}$ is implemented as the generic quasiperiodic Hamiltonian introduced in \cref{Eq: H_mpo_sum}. To complete the construction, it remains to specify an MPS element encoding that relates physical sites to tensor elements and implement the encoding-dependent shift tensor $\hat{\mathcal{K}}$.

The Hamiltonian $\hat{\mathcal{H}}_S$ is encoded using a numeration system whose basis is formed by the denominators $q_l$ of the continued-fraction convergents of the silver-mean $\alpha_{\textrm{sm}}$. We refer to this encoding as the \textit{silver-mean encoding} and to the representation of $n$ in this numeration system as its \textit{silver-mean representation}. In this encoding, each matrix element $[H_S]_{n,m}$ is mapped to a tensor element ${\mathcal{H}_S}^{\sigma\sigma'}$, where the matrix indices $n$ and $m$ are related to the tensor index sets $\sigma$ and $\sigma'$ through their silver-mean representations, $n=(\sigma_1\sigma_2\cdots\sigma_L)_S$ and $m=(\sigma'_1\sigma'_2\cdots\sigma'_L)_S$. Although the operator $\hat{\mathcal{H}}_S$ acts on the full $3^L$-dimensional Hilbert space, only those basis states corresponding to valid silver-mean representations represent physical lattice sites. Consequently, a Hamiltonian consisting of $L$ local tensors describes a physical system of size $q_L$.

It remains to construct the shift tensor $\hat{\mathcal{K}}$, which implements the map $\ket{n}\mapsto\ket{n-1}$. Here,  $\ket{n}$ is shorthand for the tensor product state $\ket{\sigma_1}\otimes\ket{\sigma_2}\otimes\cdots\otimes\ket{\sigma_L}$, where the integer $n$ is represented through its silver-mean representation  as $n=(\sigma_1\sigma_2\cdots\sigma_L)_S$. As in \cref{Sec: Sub 2.1}, the shift operator is implemented as a sum of operators. The $i$th term performs the subtraction when the least significant nonzero digit in the 
silver-mean representation of $n$ is located at position $i$.

The $i$th term is constructed in two steps. The first step is to construct an identification operator that applied to a state $\ket{n}$ returns $\ket{n}$ if and only if the least significant nonzero digit in the silver-mean  representation of $n$ is at position $i$ and annihilates the state otherwise. To this end, we define the following local operators on the 3-dimensional local Hilbert spaces: $\hat{p}^0=\ket{0}\bra{0}$, $\hat{p}^1=\ket{1}\bra{1}$, $\hat{p}^2=\ket{2}\bra{2}$ and $\hat{p}^{12}=\hat{p}^1+\hat{p}^2$. The identification operator is then given by
\begin{equation}
    I_1\otimes\cdots\otimes I_{i-1}\otimes\hat{p}_i^{12}\otimes\hat{p}^0_{i+1}\otimes\hat{p}_{i+2}^0\otimes\cdots\otimes\hat{p}_L^0.
    \label{eq: sm_identify}
\end{equation}

The second step is to construct an operator that performs the subtraction when the least significant nonzero digit is known to be at position $i$. In the digit representation, this operation corresponds to subtracting one from the least significant nonzero digit $\sigma_i$ and replacing the subsequent digits by the alternating pattern $20$ until the end of the representation is reached. In the tensor representation, these digit-wise operations are implemented by the local operators $\hat\tau^-=\ket0\bra1+\ket1\bra2$ and $\hat{r}^{02}=\ket2\bra0$. The subtraction operator is then given by
\begin{equation}
    I_1\otimes\cdots\otimes I_{i-1}\otimes\hat\tau_i^-\otimes\hat{r}^{02}_{i+1}\otimes I_{i+2}\otimes \hat{r}^{02}_{i+3}\otimes I_{i+4}\otimes\cdots
    \label{eq: sm_subtract}
\end{equation}
where the local operators at positions $i+1,\dots,L$ alternate between $\hat{r}^{02}$ and $I$. 

The final $i$th operator is constructed by combining the operators in \cref{eq: sm_identify} and \cref{eq: sm_subtract}. The resulting operator is
\begin{equation}
    I_1\otimes\cdots\otimes I_{i-1}\otimes\hat\tau_i^-\otimes\hat{r}^{02}_{i+1}\otimes\hat{p}_{i+2}^0\otimes \hat{r}_{i+3}^{02}\otimes\hat{p}^0_{i+4}\otimes\cdots
    \label{eq: sm_combined},
\end{equation}
where the local operators at positions $i+1,\dots,L$ alternate between  $\hat{r}^{02}$ and $\hat{p}^0$. 

Finally, the operator $\hat{\mathcal{K}}$ is obtained as a sum of operators of the form in \cref{eq: sm_combined}. It can be expressed concisely as
\begin{equation}
    \hat{\mathcal{K}}=\sum_{i=1}^L\hat\tau_i^-\bigotimes_{j>i}O_j,
    \quad O_j=
    \begin{cases}
    \hat p_j^0 \quad \textrm{if} \quad (j-i) \bmod 2 = 0\\
    \hat{r}_j^{02} \quad \mathrm{otherwise}
    \end{cases},
    \label{eq: K  tribo }
\end{equation}
where the identity operators acting on the remaining sites are implicit. 

Using the silver-mean encoding, the resulting off-diagonal Hamiltonian $\hat{\mathcal{H}}_S$ has a maximum bond dimension $\chi_{\textrm{max}}=8$. 

\subsubsection{Spectral properties}

We now employ the silver-mean trace discussed in App.~\ref{App: sm trace} as part of the tensorized KPM methodology, with $N_\mu = 2,300$, to compute the DOS of the silver-mean chain  with $N=q_L=54,608,393$ atoms and open boundary conditions. The hopping amplitudes follow the silver-mean word characterized by the parameters $t_A=1$ and $t_B=2$. During the Chebyshev recursion, we set the maximum bond dimension to $\chi_{\textrm{max}}=600$ and the truncation tolerance to $\textrm{tol}=10^{-6}$. The resulting DOS of the silver-mean chain is depicted in \cref{fig: sm automaton+DOS} (b)-(c). The silver-mean spectrum is symmetric around $E=0$. Furthermore, successive zoom ins into the central region reveal a self-similar structure.


\subsection{The Tribonacci quasicrystal}

\subsubsection{MPS representation}

For the non-Sturmian class, we explicitly construct the MPS representation of the Tribonacci word, which is generated through the substitution rules $A\to AB$, $B\to AC$ and $C\to A$ with initial conditions $W_0=A$, $W_1=AB$ and $W_2=ABAC$. The natural basis associated to the word is formed by the Tribonacci numbers $T_l$, defined by $T_{l}=T_{l-1}+T_{l-2}+T_{l-3}$ for $l\ge3$ with initial conditions $T_0=T_1=0$, $T_2=1$. 
Every non-negative integer $n$ can be written uniquely as a sum of Tribonacci numbers $n=\sum_{l=0}^{L-1}\sigma_{L-l}T_{l+3}$, where $\sigma_l\in\{0,1\}$ and the digits satisfy the condition $\sigma_l\sigma_{l+1}\sigma_{l+2}=0$. The $n$th symbol of the Tribonacci word, indexing from zero, can be recovered from the following rule:
\begin{enumerate}
    \item Get the last two digits of the Tribonacci representation of $n$.
    \item The $n$th symbol is $A$, $B$ or $C$ if the representation ends in 00/10, 01 or 11, respectively.
\end{enumerate}

\begin{figure}[t]
    \centering
    \includegraphics[width=\textwidth]{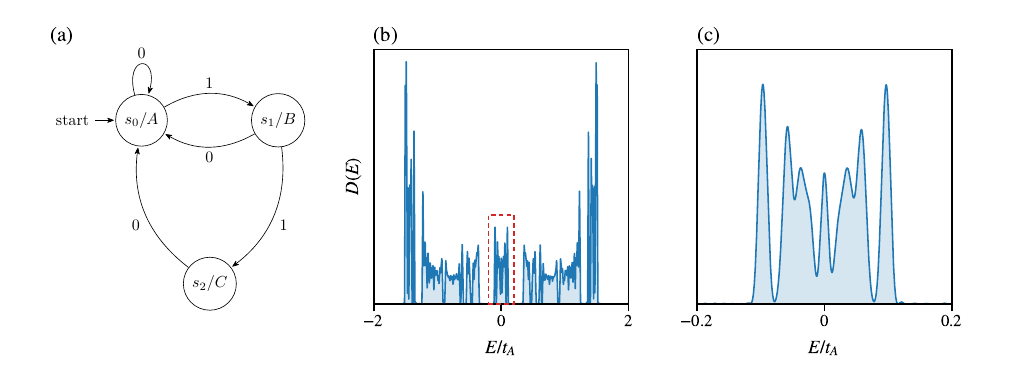}
    \caption{(a) The transition diagram of the DFAO generating the Tribonacci word. Each node is labeled as $s_i/\omega_i$, where $\omega_i\in\{A, B, C\}$ is the output associated with the final state $s_i$. Transition labeled by 1 from $s_2$ is absent because a valid Tribonacci representation cannot contain three consecutive ones. (b) DOS of the Tribonacci chain containing $N>10^6$ atoms computed using the KPM with $N_\mu=1,000$. (c) Enlarged view of the central region indicated by the red rectangle in panel (b).}
    \label{fig: tribonacci automaton+DOS}
\end{figure}

The Tribonacci word is generated by a 3-state DFAO that processes the digits of the input Tribonacci representation sequentially from the most to the least significant digit. During the process the automaton keeps track of the last two digits of the substring processed so far. The states $s_0$, $s_1$ and $s_2$ correspond to the last digits being 00/10, 01 and 11, respectively. The corresponding outputs are $A$, $B$ and $C$. The transitions between the states are depicted in \cref{fig: tribonacci automaton+DOS} (a).

The initial state of the DFAO is $s_0$. After reading the first input digit $\sigma_1$, the automaton remains in $s_0$ when $\sigma_1=0$, and transitions to $s_1$ when $\sigma_1=1$. The first local tensor is therefore given by
\begin{equation*}
    [M_1]^0=
    \begin{bmatrix}
        1 & 0 & 0
    \end{bmatrix},
    \quad
    [M_1]^1=
    \begin{bmatrix}
        0 & 1 & 0
    \end{bmatrix}.
\end{equation*}
During the tensor contraction, the current state of the automaton is represented by a one-hot vector whose $i$th component is equal to 1 if and only if the automaton is in state $s_i$. 

The local bulk tensors encode the transition rules and are given by
\begin{equation*}
    [M_l]^0=
    \begin{bmatrix}
        1 & 0 & 0 \\
        1 & 0 & 0\\
        1 & 0 & 0
    \end{bmatrix},
    \quad
    [M_l]^1=
    \begin{bmatrix}
        0 & 1 & 0\\
        0 & 0 & 1\\
        0 & 0 & 0
    \end{bmatrix},
\end{equation*}
where $1<l<L$. The vanishing third row of $[M_l]^1$ encodes the fact that Tribonacci representation cannot contain three consecutive ones. Therefore, there is no transition from state $s_2$ corresponding to input digit $\sigma_l=1$.

The final local tensor depends on the last input digit $\sigma_L$ and maps the automaton state to its corresponding output.
It is given by
\begin{equation*}
    [M_L]^0=
    \begin{bmatrix}
        A & A & A
    \end{bmatrix}^\intercal
    ,\quad
    [M_L]^1=
    \begin{bmatrix}
        B & C & 0
    \end{bmatrix}^\intercal.
\end{equation*}
With these local tensors, the $n$th element of the Tribonacci word, whose Tribonacci representation is $(\sigma_1\sigma_2\cdots\sigma_L)$,  is directly given by the tensor element $W^{\sigma_1\sigma_2\cdots\sigma_L}$. The resulting MPS has bond dimension 3, equal to the number of DFAO states, and local physical dimension 2, corresponding to the two possible numeration digits. 

\subsubsection{Tribonacci Hamiltonian} \label{App: triboancci Ham}

As with the silver-mean Hamiltonian $\hat{\mathcal{H}}_S$ in App.~\ref{App: silver-mean ham}, to complete the construction of the Tribonacci Hamiltonian $\hat{\mathcal{H}}_T$, it remains to specify the MPS element encoding relating physical sites to tensor elements and implement the encoding-dependent shift tensor $\hat{\mathcal{K}}$.

The Hamiltonian $\hat{\mathcal{H}}_T$ is encoded using a numeration system whose basis is formed by the Tribonacci numbers, referred to as the \textit{Tribonacci element encoding}. In this encoding, each matrix element $[H_T]_{n,m}$ is mapped to a tensor element ${\mathcal{H}_T}^{\sigma\sigma'}$, where the matrix indices $n$ and $m$ are related to the tensor index sets $\sigma$ and $\sigma'$ through their Tribonacci representations, $n=(\sigma_1\sigma_2\cdots\sigma_L)_T$ and $m=(\sigma'_1\sigma'_2\cdots\sigma'_L)_T$. Although the operator $\hat{\mathcal{H}}_T$ acts on the full $2^L$-dimensional Hilbert space, only the basis states corresponding to valid Tribonacci strings represent physical lattice sites. Therefore, a Hamiltonian consisting of $L$ local tensors represents a system of size equal to the $(L+3)$th Tribonacci number, denoted by $T_{L+3}$.

It remains to construct the shift tensor $\hat{\mathcal{K}}$, which implements the map $\ket{n}\mapsto\ket{n-1}$ where the Tribonacci representation of $n$ determines the local tensors of the tensor product state $\ket n$. As in \cref{Sec: Sub 2.1} and App.~\ref{App: silver-mean ham}, the shift operator is implemented as a sum of operators. The $i$th term performs the subtraction when the least significant 1 digit in the Tribonacci representation of $n$ is located at position $i$.

The $i$th term is constructed in two steps. The first step is to construct an identification operator that, applied to a state $\ket{n}$, returns $\ket{n}$ if and only if the least significant digit 1 in the Tribonacci representation of $n$ is at position $i$ and annihilates the state otherwise. The identification operator is given by
\begin{equation}
        I_1 \otimes\cdots\otimes I_{i-1}\otimes\hat p_i^1\otimes \hat p_{i+1}^0 \otimes\hat p_{i+2}^0\otimes\cdots\otimes\hat p_L^0.
        \label{eq: tribo_identify}
\end{equation}

The second step is to construct an operator that performs the subtraction when the least significant digit 1 is known to be at position $i$. The operator is given by
\begin{equation}
    I_1\otimes\cdots\otimes I_{i-1}\otimes\hat\tau_i^-\otimes\hat\tau^+_{i+1}\otimes\hat\tau_{i+2}^+\otimes I_{i+3}\otimes\hat\tau^+_{i+4}\otimes\hat\tau_{i+5}^+\otimes I_{i+6}\otimes\cdots
    \label{eq: tribo_subtract}
\end{equation}
where the local operators at positions $i+1,\dots,L$ alternate between  $I$ and $\hat\tau^+$ such that the operator at position $i+k$ is $I$ if $k \bmod 3= 0$. In the digit representation, the subtraction operation corresponds to setting the least significant digit 1 at position $i$ to 0 and letting the tail repeat the pattern 110. 

The final $i$th operator is constructed by combining the operators in \cref{eq: tribo_identify} and \cref{eq: tribo_subtract}. The resulting operator is
\begin{equation}
    I_1\otimes\cdots\otimes I_{i-1}\otimes\hat\tau_i^-\otimes\hat\tau^+_{i+1}\otimes\hat\tau_{i+2}^+\otimes \hat{p}_{i+3}^0\otimes\hat\tau^+_{i+4}\otimes\hat\tau_{i+5}^+\otimes \hat{p}_{i+6}^0\otimes\cdots
    \label{eq: tribo_combined},
\end{equation}
where the local operators at positions $i+1,\dots,L$ repeat the pattern $\hat\tau^+$, $\hat\tau^+$, $\hat{p}^0$ until the operator is constructed of $L$ local operators. Finally, the operator $\hat{\mathcal{K}}$ is obtained as a sum of operators of the form in \cref{eq: tribo_combined}. It can be expressed concisely as
\begin{equation}
    \hat{\mathcal{K}}=\sum_{i=1}^L\hat\tau_i^-\bigotimes_{j>i}O_j,
    \quad O_j=
    \begin{cases}
    \hat p_j^0 \quad \textrm{if} \quad (j-i) \bmod 3 = 0\\
    \hat\tau_j^+ \quad \mathrm{otherwise}
    \end{cases},
    \label{eq: K  tribo 2}
\end{equation}
where the identity operators acting on the remaining sites are implicit. 

Using the Tribonacci encoding, the resulting off-diagonal Hamiltonian $\hat{\mathcal{H}}_T$ has a maximum bond dimension $\chi_\textrm{max} = 12$.

\subsubsection{Spectral properties}

We now employ the Tribonacci-restricted trace discussed in App.~\ref{App: Tribo trace} as part of the tensorized KPM methodology, with $N_\mu = 1,000$, to compute the DOS of the Tribonacci chain  with $N=T_{L+3}=4,700,770$ atoms and open boundary conditions. The hopping amplitudes follow the Tribonacci word characterized by the parameters $t^{}_A$, $t^{}_B$ and $t^{}_C$, while the on-site potentials are set to zero. The Tribonacci parameters are expressed by a single control parameter $\rho\in[0,1]$ satisfying $t^{}_A/t^{}_B=t^{}_B/t^{}_C=\rho$. Smaller values of $\rho$ lead to more clearly separated main spectral bands \cite{Krebbekx2023MultifractalChains}. We set $\rho=0.8$ and $t^{}_C=1$, with $t^{}_A$ and $t^{}_B$ determined accordingly. 

During the Chebyshev recursion, we set the maximum bond dimension to $\chi_{\textrm{max}}=700$ and the truncation tolerance to $\textrm{tol}=10^{-6}$. The resulting DOS of the Tribonacci chain is depicted in \cref{fig: tribonacci automaton+DOS} (b)-(c). The Tribonacci spectrum is symmetric around $E=0$ and consists of five main clusters \cite{Niu1991SpectralStructures, Krebbekx2023MultifractalChains}. Furthermore, successive zoom ins into the central region reveal a self-similar structure.

\twocolumngrid

\bibliography{Refs,references-2}

@article{Orus2014AStates,
    title = {{A practical introduction to tensor networks: Matrix product states and projected entangled pair states}},
    year = {2014},
    journal = {Annals of Physics},
    author = {Or{\'{u}}s, Román},
    month = {10},
    pages = {117--158},
    volume = {349},
    publisher = {Elsevier BV},
    url = {http://dx.doi.org/10.1016/j.aop.2014.06.013},
    doi = {10.1016/j.aop.2014.06.013},
    issn = {0003-4916}
}

@phdthesis{Critch2013AlgebraicModels,
    title = {{Algebraic Geometry of Hidden Markov and Related Models}},
    year = {2013},
    author = {Critch, Andrew},
    school = {University of California},
    address = {Berkeley}
}

@article{Critch2014AlgebraicStates,
    title = {{Algebraic Geometry of Matrix Product States}},
    year = {2014},
    journal = {Symmetry, Integrability and Geometry: Methods and Applications},
    author = {Critch, Andrew},
    month = {9},
    pages = {095--105},
    volume = {10},
    doi = {10.3842/SIGMA.2014.095},
    issn = {18150659}
}

@article{Kocabova2007AmbiguitySystem,
    title = {{Ambiguity in the m-bonacci numeration system}},
    year = {2007},
    journal = {Discrete Mathematics {\&} Theoretical Computer Science},
    author = {Koc{\'{a}}bov{\'{a}}, Petra and Mas{\'{a}}kov{\'{a}}, Zuzana and Pelantov{\'{a}}, Edita},
    month = {1},
    volume = {9},
    number = {2},
    pages = {109--124},
    doi = {10.46298/dmtcs.381},
    issn = {1365-8050}
}

@article{Smith2024AnMonotile,
    title = {{An aperiodic monotile}},
    year = {2024},
    journal = {Comb. Theory},
    author = {Smith, David and Myers, Joseph Samuel and Kaplan, Craig S. and Goodman-Strauss, Chaim},
    number = {1},
    month = {7},
    volume = {4},
    pages = {\#6},
    doi = {10.5070/C64163843},
    issn = {2766-1334}
}

@article{Piechon1996AnomalousChains,
    title = {{Anomalous Diffusion Properties of Wave Packets on Quasiperiodic Chains}},
    year = {1996},
    journal = {Phys. Rev. Lett.},
    author = {Pi{\'{e}}chon, Frédéric},
    number = {23},
    month = {6},
    pages = {4372--4375},
    volume = {76},
    doi = {10.1103/PhysRevLett.76.4372},
    issn = {0031-9007}
}

@article{Moustaj2025AnomalousInsulators,
    title = {{Anomalous Polarization in One-Dimensional Aperiodic Insulators}},
    year = {2025},
    journal = {Condens. Matter.},
    author = {Moustaj, Anouar and Krebbekx, Julius and Morais Smith, Cristiane},
    number = {1},
    month = {1},
    pages = {3},
    volume = {10},
    doi = {10.3390/condmat10010003},
    issn = {2410-3896}
}

@article{Berthe2001AutourDOstrowski,
    title = {{Autour du syst{\`{e}}me de num{\'{e}}ration d'Ostrowski}},
    year = {2001},
    journal = {Bulletin of the Belgian Mathematical Society - Simon Stevin},
    author = {Berth{\'{e}}, Val{\'{e}}rie},
    number = {2},
    month = {1},
    volume = {8},
    pages = {209--239},
    doi = {10.36045/bbms/1102714170},
    issn = {1370-1444}
}

@article{Ostrowski1922BemerkungenApproximationen,
    title = {{Bemerkungen zur Theorie der Diophantischen Approximationen}},
    year = {1922},
    journal = {Abhandlungen aus dem Mathematischen Seminar der Universit{\"{a}}t Hamburg},
    author = {Ostrowski, Alexander},
    number = {1},
    month = {12},
    pages = {77--98},
    volume = {1},
    doi = {10.1007/BF02940581},
    issn = {0025-5858}
}

@article{Rai2021BulkQuasicrystal,
    title = {{Bulk topological signatures of a quasicrystal}},
    year = {2021},
    journal = {Phys. Rev. B},
    author = {Rai, Gautam and Schl{\"{o}}mer, Henning and Matsumura, Chris and Haas, Stephan and Jagannathan, Anuradha},
    number = {18},
    month = {11},
    pages = {184202},
    volume = {104},
    publisher = {American Physical Society},
    url = {https://link.aps.org/doi/10.1103/PhysRevB.104.184202},
    doi = {10.1103/PhysRevB.104.184202},
    issn = {2469-9950}
}

@article{Kohmoto1984CantorMap,
    title = {{Cantor spectrum for an almost periodic Schr{\"{o}}dinger equation and a dynamical map}},
    year = {1984},
    journal = {Phys. Lett. A},
    author = {Kohmoto, M. and Oono, Y.},
    number = {4},
    month = {5},
    pages = {145--148},
    volume = {102},
    doi = {10.1016/0375-9601(84)90928-9},
    issn = {03759601}
}

@article{Fishman2022CodebaseITensor,
    title = {{Codebase release 0.3 for ITensor}},
    year = {2022},
    journal = {SciPost Phys. Codebases},
    author = {Fishman, Matthew and White, Steven R. and Stoudenmire, E. Miles},
    month = {8},
    volume = {4-r0.3},
    doi = {10.21468/SciPostPhysCodeb.4-r0.3}
}

@article{Rampersad2018CommonSequences,
    title = {{Common factors in automatic and Sturmian sequences}},
    year = {2018},
    journal = {arXiv: 1802.00325},
    author = {Rampersad, Narad and Shallit, Jeffrey},
    month = {2},
    arxivId = {1802.00325}
}

@article{Takahashi2025CompactnessPropagators,
    title = {{Compactness of quantics tensor train representations of local imaginary-time propagators}},
    year = {2025},
    journal = {SciPost Phys.},
    author = {Takahashi, Haruto and Sakurai, Rihito and Shinaoka, Hiroshi},
    pages = {7},
    volume = {18},
    publisher = {SciPost},
    url = {https://scipost.org/10.21468/SciPostPhys.18.1.007}
}

@article{Jeannin2025ComprehensiveBlockade,
    title = {{Comprehensive study of out-of-equilibrium Kondo effect and Coulomb blockade}},
    year = {2025},
    journal = {Phys. Rev. B},
    author = {Jeannin, Matthieu and N{\'{u}}{\~{n}}ez-Fern{\'{a}}ndez, Yuriel and Kloss, Thomas and Parcollet, Olivier and Waintal, Xavier},
    number = {15},
    month = {10},
    volume = {112},
    pages = {155159},
    publisher = {American Physical Society (APS)},
    doi = {10.1103/9yzc-rnzh},
    issn = {2469-9969}
}

@article{Li2024ConnectingLearning,
    title = {{Connecting weighted automata, tensor networks and recurrent neural networks through spectral learning}},
    year = {2024},
    journal = {Machine Learning},
    author = {Li, Tianyu and Precup, Doina and Rabusseau, Guillaume},
    number = {5},
    month = {5},
    pages = {2619--2653},
    volume = {113},
    doi = {10.1007/s10994-022-06164-1},
    issn = {0885-6125}
}

@article{Fumega2025CorrelatedAlgorithm,
    title = {{Correlated states in super-moir{\'{e}} materials with a kernel polynomial quantics tensor cross interpolation algorithm}},
    year = {2025},
    journal = {2D Mater.},
    author = {Fumega, Adolfo O and Niedermeier, Marcel and Lado, Jose L},
    number = {1},
    month = {1},
    pages = {015018},
    volume = {12},
    doi = {10.1088/2053-1583/ad9d59},
    issn = {2053-1583}
}

@inproceedings{Baranwal2019CriticalSystem,
    title = {{Critical Exponent of Infinite Balanced Words via the Pell Number System}},
    year = {2019},
    booktitle = {Combinatorics on Words},
    author = {Baranwal, Aseem R and Shallit, Jeffrey},
    editor = {Merca{\c{s}}, Robert and Reidenbach, Daniel},
    series = {Lecture Notes in Computer Science},
    volume = {11682},
    pages = {80--92},
    publisher = {Springer International Publishing},
    address = {Cham},
    isbn = {978-3-030-28796-2}
}

@article{Kohmoto1987CriticalModel,
    title = {{Critical wave functions and a Cantor-set spectrum of a one-dimensional quasicrystal model}},
    year = {1987},
    journal = {Phys. Rev. B},
    author = {Kohmoto, Mahito and Sutherland, Bill and Tang, Chao},
    number = {3},
    month = {1},
    pages = {1020--1033},
    volume = {35},
    doi = {10.1103/PhysRevB.35.1020},
    issn = {0163-1829}
}

@article{Jeannin2024Cross-extrapolationRegime,
    title = {{Cross-extrapolation reconstruction of low-rank functions and application to quantum many-body observables in the strong coupling regime}},
    year = {2024},
    journal = {Phys. Rev. B},
    author = {Jeannin, Matthieu and N{\'{u}}{\~{n}}ez-Fern{\'{a}}ndez, Yuriel and Kloss, Thomas and Parcollet, Olivier and Waintal, Xavier},
    number = {3},
    month = {7},
    pages = {035124},
    volume = {110},
    publisher = {American Physical Society},
    url = {https://link.aps.org/doi/10.1103/PhysRevB.110.035124}
}

@article{Dilip2022DataLearning,
    title = {{Data compression for quantum machine learning}},
    year = {2022},
    journal = {Phys. Rev. Res.},
    author = {Dilip, Rohit and Liu, Yu-Jie and Smith, Adam and Pollmann, Frank},
    number = {4},
    month = {10},
    pages = {043007},
    volume = {4},
    publisher = {American Physical Society},
    url = {https://link.aps.org/doi/10.1103/PhysRevResearch.4.043007}
}

@article{Mousavi2016DecisionResults,
    title = {{Decision algorithms for Fibonacci-automatic Words, I: Basic results}},
    year = {2016},
    journal = {RAIRO - Theoretical Informatics and Applications},
    author = {Mousavi, Hamoon and Schaeffer, Luke and Shallit, Jeffrey},
    number = {1},
    month = {1},
    pages = {39--66},
    volume = {50},
    doi = {10.1051/ita/2016010},
    issn = {0988-3754}
}

@article{White1992DensityGroups,
    title = {{Density matrix formulation for quantum renormalization groups}},
    year = {1992},
    journal = {Phys. Rev. Lett.},
    author = {White, Steven R.},
    number = {19},
    month = {11},
    pages = {2863--2866},
    volume = {69},
    doi = {10.1103/PhysRevLett.69.2863},
    issn = {0031-9007}
}

@article{Moustaj2021EffectsQuasicrystal,
    title = {{Effects of disorder in the Fibonacci quasicrystal}},
    year = {2021},
    journal = {Phys. Rev. B},
    author = {Moustaj, A. and Kempkes, S. and Morais Smith, C.},
    number = {14},
    pages = {144201},
    volume = {104},
    url = {https://doi.org/10.1103/PhysRevB.104.144201},
    issn = {24699969}
}

@article{Sun2024EnhancementChain,
    title = {{Enhancement of superconductivity in the Fibonacci chain}},
    year = {2024},
    journal = {Phys. Rev. B},
    author = {Sun, Meng and {\v{C}}ade{\v{z}}, Tilen and Yurkevich, Igor and Andreanov, Alexei},
    number = {13},
    month = {4},
    pages = {134504},
    volume = {109},
    doi = {10.1103/PhysRevB.109.134504},
    issn = {2469-9950}
}

@article{Sire1990ExcitationQuasicrystals,
    title = {{Excitation spectrum, extended states, gap closing: some exact results for codimension one quasicrystals}},
    year = {1990},
    journal = {J. Phys. France},
    author = {Sire, C and Mosseri, R},
    pages = {1569--1583},
    volume = {51},
    doi = {10.1051/jphys:0199000510150156900}
}

@article{RocheCarrasco2025FamilyTensor,
    title = {{Family of Aperiodic Tilings with Tunable Quantum Geometric Tensor}},
    year = {2025},
    journal = {Phys. Rev. Lett.},
    author = {Roche Carrasco, Hector and Schirmann, Justin and Mordret, Aurelien and Grushin, Adolfo G.},
    number = {23},
    month = {12},
    pages = {236603},
    volume = {135},
    doi = {10.1103/dzqm-9kwj},
    issn = {0031-9007}
}

@article{Carlitz1972FibonacciOrder,
    title = {{Fibonacci Representations of Higher Order}},
    year = {1972},
    journal = {The Fibonacci Quarterly},
    author = {Carlitz, L. and Scoville, Richard and Hoggatt, V.E.},
    number = {1},
    month = {1},
    pages = {43--70},
    volume = {10},
    doi = {10.1080/00150517.1972.12430969},
    issn = {0015-0517}
}

@incollection{Allouche2003FiniteComputation,
    title = {{Finite Automata and Other Models of Computation}},
    year = {2003},
    booktitle = {Automatic Sequences: Theory, Applications, Generalizations},
    author = {Allouche, Jean-Paul and Shallit, Jeffrey},
    chapter = {4},
    pages = {128--151},
    publisher = {Cambridge University Press}
}

@article{Crosswhite2008FiniteAlgorithms,
    title = {{Finite automata for caching in matrix product algorithms}},
    year = {2008},
    journal = {Phys. Rev. A},
    author = {Crosswhite, Gregory M. and Bacon, Dave},
    number = {1},
    month = {7},
    pages = {012356},
    volume = {78},
    doi = {10.1103/PhysRevA.78.012356},
    issn = {1050-2947}
}

@article{Thiem2011GeneralizedSystems,
    title = {{Generalized inverse participation numbers in metallic-mean quasiperiodic systems}},
    year = {2011},
    journal = {The European Physical Journal B},
    author = {Thiem, S. and Schreiber, M.},
    number = {4},
    month = {10},
    pages = {415--421},
    volume = {83},
    doi = {10.1140/epjb/e2011-20323-7},
    issn = {1434-6028}
}

@article{Sardinero2026JosephsonStates,
    title = {{Josephson Effect in Fibonacci Superconductors from Topological Supragap States}},
    year = {2026},
    journal = {Phys. Rev. Lett.},
    author = {Sardinero, Ignacio and Cayao, Jorge and Yada, Keiji and Tanaka, Yukio and Burset, Pablo},
    number = {18},
    month = {5},
    pages = {186002},
    volume = {136},
    doi = {10.1103/6m6b-y3yv},
    issn = {0031-9007}
}

@article{NunezFernandez2022LearningTrains,
    title = {{Learning Feynman Diagrams with Tensor Trains}},
    year = {2022},
    journal = {Phys. Rev. X},
    author = {N{\'{u}}{\~{n}}ez Fern{\'{a}}ndez, Yuriel and Jeannin, Matthieu and Dumitrescu, Philipp T and Kloss, Thomas and Kaye, Jason and Parcollet, Olivier and Waintal, Xavier},
    number = {4},
    month = {11},
    pages = {041018},
    volume = {12},
    publisher = {American Physical Society},
    url = {https://link.aps.org/doi/10.1103/PhysRevX.12.041018}
}

@article{Rabusseau2014LearningDecompositions,
    title = {{Learning Negative Mixture Models by Tensor Decompositions}},
    year = {2014},
    journal = {arXiv: 1403.4224},
    author = {Rabusseau, Guillaume and Denis, François}
}

@article{NunezFernandez2025LearningLibraries,
    title = {{Learning tensor networks with tensor cross interpolation: New algorithms and libraries}},
    year = {2025},
    journal = {SciPost Phys.},
    author = {N{\'{u}}{\~{n}}ez Fern{\'{a}}ndez, Yuriel and Ritter, Marc K and Jeannin, Matthieu and Li, Jheng-Wei and Kloss, Thomas and Louvet, Thibaud and Terasaki, Satoshi and Parcollet, Olivier and von Delft, Jan and Shinaoka, Hiroshi and Waintal, Xavier},
    number = {3},
    month = {3},
    pages = {104},
    volume = {18},
    doi = {10.21468/SciPostPhys.18.3.104},
    issn = {2542-4653}
}

@article{Mace2019Many-bodyChain,
    title = {{Many-body localization in a quasiperiodic Fibonacci chain}},
    year = {2019},
    journal = {SciPost Phys.},
    author = {Mac{\'{e}}, Nicolas and Laflorencie, Nicolas and Alet, Fabien},
    number = {4},
    month = {4},
    pages = {050},
    volume = {6},
    doi = {10.21468/SciPostPhys.6.4.050},
    issn = {2542-4653}
}

@article{Chan2016MatrixAlgorithms,
    title = {{Matrix product operators,  matrix product states,  and ab initio density matrix renormalization group algorithms}},
    year = {2016},
    journal = {J. Chem. Phys.},
    author = {Chan, Garnet Kin-Lic and Keselman, Anna and Nakatani, Naoki and Li, Zhendong and White, Steven R},
    number = {1},
    month = {7},
    pages = {014102},
    volume = {145},
    publisher = {AIP Publishing},
    url = {http://dx.doi.org/10.1063/1.4955108},
    issn = {1089-7690}
}

@article{Cirac2021MatrixTheorems,
    title = {{Matrix product states and projected entangled pair states: Concepts, symmetries, theorems}},
    year = {2021},
    journal = {Rev. Mod. Phys.},
    author = {Cirac, J. Ignacio and P{\'{e}}rez-Garc{\'{i}}a, David and Schuch, Norbert and Verstraete, Frank},
    number = {4},
    month = {12},
    pages = {045003},
    volume = {93},
    doi = {10.1103/RevModPhys.93.045003},
    issn = {0034-6861}
}

@article{Shechtman1984MetallicSymmetry,
    title = {{Metallic Phase with Long-Range Orientational Order and No Translational Symmetry}},
    year = {1984},
    journal = {Phys. Rev. Lett.},
    author = {Shechtman, D and Blech, I and Gratias, D and Cahn, J W},
    number = {20},
    pages = {1951--1953},
    volume = {53},
    publisher = {American Physical Society},
    url = {https://link.aps.org/doi/10.1103/PhysRevLett.53.1951},
    doi = {10.1103/PhysRevLett.53.1951}
}

@article{Krebbekx2023MultifractalChains,
    title = {{Multifractal properties of Tribonacci chains}},
    year = {2023},
    journal = {Phys. Rev. B},
    author = {Krebbekx, J. P. J. and Moustaj, A. and Dajani, K. and Morais Smith, C.},
    number = {10},
    month = {9},
    pages = {104204},
    volume = {108},
    doi = {10.1103/PhysRevB.108.104204},
    issn = {2469-9950}
}

@article{Shinaoka2023MultiscaleTrains,
    title = {{Multiscale Space-Time Ansatz for Correlation Functions of Quantum Systems Based on Quantics Tensor Trains}},
    year = {2023},
    journal = {Phys. Rev. X},
    author = {Shinaoka, Hiroshi and Wallerberger, Markus and Murakami, Yuta and Nogaki, Kosuke and Sakurai, Rihito and Werner, Philipp and Kauch, Anna},
    number = {2},
    month = {4},
    pages = {021015},
    volume = {13},
    publisher = {American Physical Society},
    url = {https://link.aps.org/doi/10.1103/PhysRevX.13.021015}
}

@article{Rauzy1982NombresSubstitutions,
    title = {{Nombres alg{\'{e}}briques et substitutions}},
    year = {1982},
    journal = {Bull. Soc. Math. France},
    author = {Rauzy, G{\'{e}}rard},
    pages = {147--178},
    volume = {110},
    doi = {10.24033/bsmf.1957},
    issn = {0037-9484}
}

@article{Jagannathan2019NonmonotonicQuasicrystal,
    title = {{Nonmonotonic crossover and scaling behavior in a disordered one-dimensional quasicrystal}},
    year = {2019},
    journal = {Phys. Rev. B},
    author = {Jagannathan, Anuradha and Jeena, Piyush and Tarzia, Marco},
    number = {5},
    month = {2},
    pages = {054203},
    volume = {99},
    publisher = {American Physical Society},
    url = {https://link.aps.org/doi/10.1103/PhysRevB.99.054203},
    doi = {10.1103/PhysRevB.99.054203}
}

@incollection{Frougny2010NumberAutomata,
    title = {{Number representation and finite automata}},
    year = {2010},
    booktitle = {Combinatorics, Automata and Number Theory},
    author = {Frougny, CH. and Sakarovitch, J.},
    editor = {Berth{\'{e}}, Valerie and Rigo, Michel},
    chapter = {2},
    pages = {34--107},
    publisher = {Cambridge University Press}
}

@article{Jackson1912OnPolynomials,
    title = {{On approximation by trigonometric sums and polynomials}},
    year = {1912},
    journal = {Transactions of the American Mathematical Society},
    author = {Jackson, Dunham},
    number = {4},
    pages = {491--515},
    volume = {13},
    publisher = {American Mathematical Society (AMS)},
    url = {http://dx.doi.org/10.1090/S0002-9947-1912-1500930-2},
    doi = {10.1090/s0002-9947-1912-1500930-2},
    issn = {1088-6850}
}

@article{Hieronymi2018OstrowskiAutomata,
    title = {{Ostrowski Numeration Systems, Addition, and Finite Automata}},
    year = {2018},
    journal = {Notre Dame Journal of Formal Logic},
    author = {Hieronymi, Philipp and Terry Jr., Alonza},
    number = {2},
    month = {1},
    volume = {59},
    pages = {215--232},
    doi = {10.1215/00294527-2017-0027},
    issn = {0029-4527}
}

@article{Schirmann2024PhysicalModes,
    title = {{Physical Properties of an Aperiodic Monotile with Graphene-like Features, Chirality, and Zero Modes}},
    year = {2024},
    journal = {Phys. Rev. Lett.},
    author = {Schirmann, Justin and Franca, Selma and Flicker, Felix and Grushin, Adolfo G.},
    number = {8},
    month = {2},
    pages = {086402},
    volume = {132},
    doi = {10.1103/PhysRevLett.132.086402},
    issn = {0031-9007}
}

@article{Rai2019ProximityRing,
    title = {{Proximity effect in a superconductor-quasicrystal hybrid ring}},
    year = {2019},
    journal = {Phys. Rev. B},
    author = {Rai, Gautam and Haas, Stephan and Jagannathan, Anuradha},
    number = {16},
    month = {10},
    pages = {165121},
    volume = {100},
    publisher = {American Physical Society},
    doi = {10.1103/PhysRevB.100.165121}
}

@article{Ritter2024QuanticsFunctions,
    title = {{Quantics Tensor Cross Interpolation for High-Resolution Parsimonious Representations of Multivariate Functions}},
    year = {2024},
    journal = {Phys. Rev. Lett.},
    author = {Ritter, Marc K. and N{\'{u}}{\~{n}}ez Fern{\'{a}}ndez, Yuriel and Wallerberger, Markus and von Delft, Jan and Shinaoka, Hiroshi and Waintal, Xavier},
    number = {5},
    month = {1},
    pages = {056501},
    volume = {132},
    doi = {10.1103/PhysRevLett.132.056501},
    issn = {0031-9007}
}

@misc{Ritter2022QuanticsTCI.jl,
    title = {{QuanticsTCI.jl}},
    year = {2022},
    author = {Ritter, Marc and {contributors}},
    month = {7},
    url = {https://github.com/tensor4all/quanticstci.jl}
}

@article{Sun2026Real-spaceNetworks,
    title = {{Real-space spectral functions of three-dimensional billion-size topological non-Hermitian matter with tensor networks}},
    year = {2026},
    journal = {arXiv:2606.16424},
    author = {Sun, Yitao and Lado, Jose L. and Chen, Guangze},
    month = {6},
    arxivId = {2606.16424}
}

@article{Niu1986Renormalization-GroupSystems,
    title = {{Renormalization-Group Study of One-Dimensional Quasiperiodic Systems}},
    year = {1986},
    journal = {Phys. Rev. Lett.},
    author = {Niu, Qian and Nori, Franco},
    number = {16},
    month = {10},
    pages = {2057--2060},
    volume = {57},
    doi = {10.1103/PhysRevLett.57.2057},
    issn = {0031-9007}
}

@article{Zeckendorf1972RepresentationsLucas,
    title = {{Repr{\'{e}}sentation des nombres naturels par une somme de nombres de Fibonacci ou de nombres de Lucas}},
    year = {1972},
    journal = {Bulletin de La Soci{\'{e}}t{\'{e}} Royale des Sciences de Li{\`{e}}ge},
    author = {Zeckendorf, {\'{E}}douard},
    pages = {179--182},
    volume = {41}
}

@article{Baake2019ScalingMeasure,
    title = {{Scaling properties of the thue–morse measure}},
    year = {2019},
    journal = {Discrete Contin. Dyn. Syst.},
    author = {Baake, M. and Gohlke, P. and Kesseb{\"{o}}hmer, M. and Schindler, T.},
    number = {7},
    pages = {4157--4185},
    volume = {39},
    doi = {10.3934/dcds.2019168},
    issn = {1553-5231}
}

@article{Sun2025Self-consistentSites,
    title = {{Self-consistent tensor network method for correlated super-moir{\'{e}} matter beyond one billion sites}},
    year = {2025},
    journal = {Phys. Rev. Res.},
    author = {Sun, Yitao and Niedermeier, Marcel and Ant{\~{a}}o, Tiago V. C. and Fumega, Adolfo O. and Lado, Jose L.},
    number = {4},
    month = {12},
    pages = {043288},
    volume = {7},
    doi = {10.1103/krjp-mn4v},
    issn = {2643-1564}
}

@article{Rudin1959SomeCoefficients,
    title = {{Some theorems on Fourier coefficients}},
    year = {1959},
    journal = {Proceedings of the American Mathematical Society},
    author = {Rudin, Walter},
    number = {6},
    month = {12},
    pages = {855--859},
    volume = {10},
    doi = {10.1090/S0002-9939-1959-0116184-5},
    issn = {1088-6826}
}

@article{Cerovski2005SpectralDimensions,
    title = {{Spectral and diffusive properties of silver-mean quasicrystals in one, two, and three dimensions}},
    year = {2005},
    journal = {Phys. Rev. B},
    author = {Cerovski, V. Z. and Schreiber, M. and Grimm, U.},
    number = {5},
    month = {8},
    pages = {054203},
    volume = {72},
    doi = {10.1103/PhysRevB.72.054203},
    issn = {1098-0121}
}

@article{Niu1991SpectralStructures,
    title = {{Spectral splitting and wave-function scaling in quasicrystalline and hierarchical structures}},
    year = {1990},
    journal = {Phys. Rev. B},
    author = {Niu, Q and Nori, F},
    pages = {10329--10341},
    volume = {42},
    doi = {10.1103/PhysRevB.42.10329}
}

@article{Stoudenmire2012StudyingGroup,
    title = {{Studying Two-Dimensional Systems with the Density Matrix Renormalization Group}},
    year = {2012},
    journal = {Annual Review of Condensed Matter Physics},
    author = {Stoudenmire, E M and White, Steven R},
    number = {1},
    month = {3},
    pages = {111--128},
    volume = {3},
    publisher = {Annual Reviews},
    url = {http://dx.doi.org/10.1146/annurev-conmatphys-020911-125018},
    doi = {10.1146/annurev-conmatphys-020911-125018},
    issn = {1947-5462}
}

@incollection{Lothaire2002SturmianWords,
    title = {{Sturmian Words}},
    year = {2002},
    booktitle = {Algebraic Combinatorics on Words},
    author = {Lothaire, M.},
    chapter = {2},
    pages = {45--110},
    publisher = {Cambridge University Press},
    address = {Cambridge}
}

@incollection{Berthe2010SubstitutionsTilings,
    title = {{Substitutions, Rauzy fractals and tilings}},
    year = {2010},
    booktitle = {Combinatorics, Automata and Number Theory},
    author = {Berth{\'{e}}, Val{\'{e}}rie and Siegel, Anne and Thuswaldner, J{\"{o}}rg},
    editor = {Berth{\'{e}}, Val{\'{e}}rie and Rigo, Michel},
    series = {Encyclopedia of Mathematics and its Applications},
    volume = {135},
    chapter = {5},
    pages = {248--323},
    publisher = {Cambridge University Press},
    address = {Cambridge}
}

@inproceedings{Stoudenmire2016SupervisedNetworks,
    title = {{Supervised Learning with Tensor Networks}},
    year = {2016},
    booktitle = {Advances in Neural Information Processing Systems},
    author = {Stoudenmire, Edwin and Schwab, David J},
    editor = {Lee, D and Sugiyama, M and Luxburg, U and Guyon, I and Garnett, R},
    pages = {},
    volume = {29},
    publisher = {Curran Associates, Inc.},
    url = {https://proceedings.neurips.cc/paper_files/paper/2016/file/5314b9674c86e3f9d1ba25ef9bb32895-Paper.pdf}
}

@article{Banuls2023TensorMap,
    title = {{Tensor Network Algorithms: A Route Map}},
    year = {2023},
    journal = {Annual Review of Condensed Matter Physics},
    author = {Ba{\~{n}}uls, Mari Carmen},
    number = {1},
    month = {3},
    pages = {173--191},
    volume = {14},
    publisher = {Annual Reviews},
    url = {http://dx.doi.org/10.1146/annurev-conmatphys-040721-022705},
    doi = {10.1146/annurev-conmatphys-040721-022705},
    issn = {1947-5462}
}

@article{Moustaj2026TensorSystems,
    title = {{Tensor network approach to momentum-resolved spectroscopy in nonperiodic super-moir{\'{e}} systems}},
    year = {2026},
    journal = {Phys. Rev. Res.},
    author = {Moustaj, Anouar and Sun, Yitao and Ant{\~{a}}o, Tiago V. C. and Lado, Jose L.},
    month = {6},
    volume = {8},
    pages = {023282},
    doi = {10.1103/9btt-y8sh},
    issn = {2643-1564}
}

@article{Antao2025TensorMosaics,
    title = {{Tensor network method for real-space topology in quasicrystal Chern mosaics}},
    year = {2025},
    journal = {arXiv: 2506.05230},
    author = {Ant{\~{a}}o, Tiago V. C. and Sun, Yitao and Fumega, Adolfo O. and Lado, Jose L.},
    month = {6},
    url = {https://arxiv.org/abs/2506.05230},
    arxivId = {2506.05230}
}

@article{Antao2026TensorApplications,
    title = {{Tensor network solvers for ultra-large tight-binding Hamiltonians: algorithms and applications}},
    year = {2026},
    journal = {arXiv:  2607.00991},
    author = {Ant{\~{a}}o, Tiago V. C. and Moustaj, Anouar and Sun, Yitao and Lado, Jose L.},
    month = {7},
    url = {https://arxiv.org/abs/2607.00991},
    arxivId = {2607.00991}
}

@article{Orus2019TensorSystems,
    title = {{Tensor networks for complex quantum systems}},
    year = {2019},
    journal = {Nature Reviews Physics},
    author = {Or{\'{u}}s, Román},
    number = {9},
    month = {8},
    pages = {538--550},
    volume = {1},
    publisher = {Springer Science and Business Media LLC},
    url = {http://dx.doi.org/10.1038/s42254-019-0086-7},
    issn = {2522-5820}
}

@article{Erpenbeck2023TensorModels,
    title = {{Tensor train continuous time solver for quantum impurity models}},
    year = {2023},
    journal = {Phys. Rev. B},
    author = {Erpenbeck, A and Lin, W.-T. and Blommel, T and Zhang, L and Iskakov, S and Bernheimer, L and N{\'{u}}{\~{n}}ez-Fern{\'{a}}ndez, Y and Cohen, G and Parcollet, O and Waintal, X and Gull, E},
    number = {24},
    month = {6},
    pages = {245135},
    volume = {107},
    publisher = {American Physical Society},
    url = {https://link.aps.org/doi/10.1103/PhysRevB.107.245135}
}

@article{Moustaj2026Tensor-networkSites,
    title = {{Tensor-network methodology for super-moir{\'{e}} excitons beyond one billion sites}},
    year = {2026},
    journal = {arXiv: 2603.02011},
    author = {Moustaj, Anouar and Sun, Yitao and Ant{\~{a}}o, Tiago V. C. and Eek, Lumen and Lado, Jose L.},
    month = {3},
    arxivId = {2603.02011}
}

@article{Oseledets2011Tensor-TrainDecomposition,
    title = {{Tensor-Train Decomposition}},
    year = {2011},
    journal = {SIAM Journal on Scientific Computing},
    author = {Oseledets, I V},
    number = {5},
    month = {1},
    pages = {2295--2317},
    volume = {33},
    publisher = {Society for Industrial {\&} Applied Mathematics (SIAM)},
    url = {http://dx.doi.org/10.1137/090752286},
    issn = {1095-7197}
}

@misc{Ritter2022TensorCrossInterpolation.jl,
    title = {{TensorCrossInterpolation.jl}},
    year = {2022},
    author = {Ritter, Marc and {contributors}},
    month = {7},
    url = {https://github.com/tensor4all/TensorCrossInterpolation.jl/tree/main}
}

@article{Jolly2025TensorizedChemistry,
    title = {{Tensorized orbitals for computational chemistry}},
    year = {2025},
    journal = {Phys. Rev. B},
    author = {Jolly, Nicolas and Fern{\'{a}}ndez, Yuriel Núñez and Waintal, Xavier},
    number = {24},
    month = {6},
    pages = {245115},
    volume = {111},
    doi = {10.1103/PhysRevB.111.245115},
    issn = {2469-9950}
}

@article{Schollwock2011TheStates,
    title = {{The density-matrix renormalization group in the age of matrix product states}},
    year = {2011},
    journal = {Annals of Physics},
    author = {Schollw{\"{o}}ck, Ulrich},
    number = {1},
    month = {1},
    pages = {96--192},
    volume = {326},
    doi = {10.1016/j.aop.2010.09.012},
    issn = {00034916}
}

@article{Jagannathan2021TheMultifractality,
    title = {{The Fibonacci quasicrystal: Case study of hidden dimensions and multifractality}},
    year = {2021},
    journal = {Rev. Mod. Phys.},
    author = {Jagannathan, Anuradha},
    number = {4},
    month = {12},
    pages = {045001},
    volume = {93},
    publisher = {American Physical Society},
    url = {https://doi.org/10.1103/RevModPhys.93.045001},
    issn = {15390756},
    arxivId = {2012.14744}
}

@article{Fishman2022TheCalculations,
    title = {{The ITensor Software Library for Tensor Network Calculations}},
    year = {2022},
    journal = {SciPost Phys. Codebases},
    author = {Fishman, Matthew and White, Steven R. and Stoudenmire, E. Miles},
    month = {8},
    volume = {4},
    doi = {10.21468/SciPostPhysCodeb.4}
}

@article{Weie2006TheMethod,
    title = {{The kernel polynomial method}},
    year = {2006},
    journal = {Rev. Mod. Phys.},
    author = {Wei{\ss}e, Alexander and Wellein, Gerhard and Alvermann, Andreas and Fehske, Holger},
    number = {1},
    month = {3},
    pages = {275--306},
    volume = {78},
    doi = {10.1103/RevModPhys.78.275},
    issn = {0034-6861}
}

@article{deSpinadel1999TheSpectra,
    title = {{The metallic means family and multifractal spectra}},
    year = {1999},
    journal = {Nonlinear Analysis: Theory, Methods {\&} Applications},
    author = {de Spinadel, Vera W.},
    number = {6},
    month = {6},
    pages = {721--745},
    volume = {36},
    doi = {10.1016/S0362-546X(98)00123-0},
    issn = {0362546X}
}

@incollection{Allouche1999TheSequence,
    title = {{The Ubiquitous Prouhet-Thue-Morse Sequence}},
    year = {1999},
    booktitle = {Sequences and their Applications},
    author = {Allouche, Jean-Paul and Shallit, Jeffrey},
    pages = {1--16},
    publisher = {Springer London},
    address = {London},
    doi = {10.1007/978-1-4471-0551-0{\_}1}
}

@article{Yoshii2021TopologicalIndex,
    title = {{Topological charge pumping in quasiperiodic systems characterized by the Bott index}},
    year = {2021},
    journal = {Phys. Rev. B},
    author = {Yoshii, Mao and Kitamura, Sota and Morimoto, Takahiro},
    number = {15},
    month = {10},
    pages = {155126},
    volume = {104},
    publisher = {American Physical Society},
    url = {https://doi.org/10.1103/PhysRevB.104.155126},
    issn = {24699969},
    arxivId = {2105.05654}
}

@article{Verbin2015TopologicalQuasicrystal,
    title = {{Topological pumping over a photonic Fibonacci quasicrystal}},
    year = {2015},
    journal = {Phys. Rev. B},
    author = {Verbin, Mor and Zilberberg, Oded and Lahini, Yoav and Kraus, Yaacov E. and Silberberg, Yaron},
    number = {6},
    month = {2},
    pages = {064201},
    volume = {91},
    doi = {10.1103/PhysRevB.91.064201},
    issn = {1098-0121}
}

@article{Kraus2012TopologicalQuasicrystals,
    title = {{Topological States and Adiabatic Pumping in Quasicrystals}},
    year = {2012},
    journal = {Phys. Rev. Lett.},
    author = {Kraus, Yaacov E and Lahini, Yoav and Ringel, Zohar and Verbin, Mor and Zilberberg, Oded},
    number = {10},
    month = {9},
    pages = {106402},
    volume = {109},
    publisher = {American Physical Society},
    url = {https://link.aps.org/doi/10.1103/PhysRevLett.109.106402},
    doi = {10.1103/PhysRevLett.109.106402}
}

@article{Huggins2019TowardsNetworks,
    title = {{Towards quantum machine learning with tensor networks}},
    year = {2019},
    journal = {Quantum Science and Technology},
    author = {Huggins, William and Patil, Piyush and Mitchell, Bradley and Whaley, K Birgitta and Stoudenmire, E Miles},
    number = {2},
    month = {1},
    pages = {024001},
    volume = {4},
    publisher = {IOP Publishing},
    url = {http://dx.doi.org/10.1088/2058-9565/aaea94},
    doi = {10.1088/2058-9565/aaea94},
    issn = {2058-9565}
}

@article{Dulea1992Trace-mapModel,
    title = {{Trace-map invariant and zero-energy states of the tight-binding Rudin-Shapiro model}},
    year = {1992},
    journal = {Phys. Rev. B},
    author = {Dulea, Mihnea and Johansson, Magnus and Riklund, Rolf},
    number = {6},
    month = {8},
    pages = {3296--3304},
    volume = {46},
    doi = {10.1103/PhysRevB.46.3296},
    issn = {0163-1829}
}

@article{Oseledets2010TT-crossArrays,
    title = {{TT-cross approximation for multidimensional arrays}},
    year = {2010},
    journal = {Linear Algebra and its Applications},
    author = {Oseledets, Ivan and Tyrtyshnikov, Eugene},
    number = {1},
    month = {1},
    pages = {70--88},
    volume = {432},
    publisher = {Elsevier BV},
    url = {http://dx.doi.org/10.1016/j.laa.2009.07.024},
    issn = {0024-3795}
}

@article{Rohshap2025Two-particleEquations,
    title = {{Two-particle calculations with quantics tensor trains: Solving the parquet equations}},
    year = {2025},
    journal = {Phys. Rev. Res.},
    author = {Rohshap, Stefan and Ritter, Marc K. and Shinaoka, Hiroshi and von Delft, Jan and Wallerberger, Markus and Kauch, Anna},
    number = {2},
    month = {4},
    pages = {023087},
    volume = {7},
    doi = {10.1103/PhysRevResearch.7.023087},
    issn = {2643-1564}
}

@article{Han2018UnsupervisedStates,
    title = {{Unsupervised Generative Modeling Using Matrix Product States}},
    year = {2018},
    journal = {Phys. Rev. X},
    author = {Han, Zhao-Yu and Wang, Jun and Fan, Heng and Wang, Lei and Zhang, Pan},
    number = {3},
    month = {7},
    pages = {031012},
    volume = {8},
    publisher = {American Physical Society},
    url = {https://link.aps.org/doi/10.1103/PhysRevX.8.031012}
}

@article{Zhou2020WhatComputers,
    title = {{What Limits the Simulation of Quantum Computers?}},
    year = {2020},
    journal = {Phys. Rev. X},
    author = {Zhou, Yiqing and Stoudenmire, E Miles and Waintal, Xavier},
    number = {4},
    month = {11},
    pages = {041038},
    volume = {10},
    publisher = {American Physical Society},
    url = {https://link.aps.org/doi/10.1103/PhysRevX.10.041038},
    doi = {10.1103/PhysRevX.10.041038}
}

@article{Waintal2026WhoTechniques,
    title = {{Who can compete with quantum computers? Lecture notes on quantum inspired tensor networks computational techniques}},
    year = {2026},
    journal = {arXiv: 2601.03035},
    author = {Waintal, Xavier and Huang, Chen-How and Groth, Christoph W},
    month = {1},
    pages = {},
    arxivId = {2601.03035}
}

@misc{millarepo,
    author = {Kolehmainen, Milla},
    title = {Tensor{Q}uasicrystals},
    howpublished = {\url{https://github.com/MillaSigrid/TensorQuasicrystals}},
    year = {2026}
}

\end{document}